\documentclass{article}

\usepackage{arxiv}

\usepackage[utf8]{inputenc} % allow utf-8 input
\usepackage[T1]{fontenc}    % use 8-bit T1 fonts
\usepackage{hyperref}       % hyperlinks
\usepackage{url}            % simple URL typesetting
\usepackage{booktabs}       % professional-quality tables
\usepackage{amsfonts}       % blackboard math symbols
\usepackage{nicefrac}       % compact symbols for 1/2, etc.
\usepackage{microtype}      % microtypography
\usepackage{lipsum}
\usepackage{xcolor,soul,framed} %,caption

\colorlet{shadecolor}{yellow}
\usepackage[pdftex]{graphicx}
\graphicspath{{../pdf/}{../jpeg/}}
\DeclareGraphicsExtensions{.pdf,.jpeg,.png}

\usepackage[cmex10]{amsmath}
\usepackage{array}
\usepackage{mdwmath}
\usepackage{mdwtab}
\usepackage{eqparbox}
\usepackage{url}
\usepackage{float}
\title{Voltage Regulator Assisted Lightweight Countermeasure Against Fault Injection Attacks}

\author{
  Ali~Vosoughi\thanks{Ali Vosoughi and Sel\c{c}uk K\"{o}se are with Electrical and Computer Engineering Department at University of Rochester, Rochester, NY 14627 USA (e-mail: mvosough@ur.rochester.edu,~selcuk.kose@rochester.edu).}\thanks{Manuscript received \today. This paper is an extension from the previously published paper from the ACM GLSVLSI conference held on May 9-11, 2019 in Washington, D.C., US~\cite{vosoughi2019leveraging}.} \\
  Department of Electrical and Computer Engineering\\
  University of Rochester\\
  Rochester, NY 14620 \\
  \texttt{mvosough@ur.rochester.edu} \\
   \And
  Longfei Wang\\
  Department of Electrical and Computer Engineering\\
  University of Rochester\\
  Rochester, NY 14620 \\
  \texttt{longfei.wang@rochester.edu} \\
   \And
  Sel\c{c}uk K\"{o}se\\
  Department of Electrical and Computer Engineering\\
  University of Rochester\\
  Rochester, NY 14620 \\
  \texttt{selcuk.kose@rochester.edu} \\
}

\begin{document}
\maketitle

\begin{abstract}
The impeccable design of sensitive and cryptographic circuits (CC) against fault injection attacks is essential for modern data storage, communication, and computation systems that are susceptible to fault injection attacks.
The robustness of a CC against voltage glitch attacks increases with an on-chip voltage regulator that considers the impact of topology and component selection on the fault injection robustness. 
With an emphasis on increasing the number of phases in a multiphase voltage regulator and component selection, an on-chip voltage regulator with multiphase configuration and an optimal operating point is proposed as a lightweight countermeasure to minimize injected glitches. 
Furthermore, an infective countermeasure is added to the on-chip multiphase voltage regulator that contaminates the computation of the cryptographic algorithm when a voltage glitch reaches to the CC. 
By increasing the number of phases from 1 to 16, the confrontation with fault attacks increases by 52.45\%, which is equal to 91.82\%  if the number of phases increases to 32. 
Using the infective countermeasure for fault resiliency, the security-enhanced CC provides a robust and resilient solution against fault attacks that improve the security and availability of the device.
\end{abstract}

% keywords can be removed
\keywords{voltage regulator\and multiphase voltage regulator \and fault injection attack \and side-channel analysis \and countermeasures \and power glitches \and infective computation \and fault resilient computation}

{S}{ide-channel} attacks are a class of cryptoanalysis attacks on the implementation of a cryptographic circuit (CC) that threaten the security of modern storage, communication, and encryption systems.
In a side-channel attack, side-channel leakages of the physically accessible CC is used to obtain the secret key of the device. 
Despite the mathematically sound schemes of the cryptographic algorithms, the presence of a side-channel attack shudders the security of these devices~\cite{skorobogatov2016bumpy, ali_iscas2019}.

Active side-channel attacks are a class of side-channel attacks that obtain the correct key by composing transient (or
permanent) abnormalities on the CC and investigating the output of CC under the attack.
In these attacks that are known as fault injection attacks, deliberate fault in the CC that an attacker creates are exploited in a fault analysis tools, such as differential fault analysis (DFA) \cite{moradi2006generalized, piret2003differential, giraud2004dfa, dusart2003differential}, safe error analysis (SEA) \cite{yen2000checking}, and collision fault analysis (CFA) \cite{blomer2006fault} to obtain the key.

Generating faulty outputs on the CC requires a different level of equipment and experience of mounting an attack that varies from expensive laser techniques to affordable clock and voltage glitch injection attacks \cite{barenghi2012fault, bar2006sorcerer}.
Each of the fault injection techniques, such as over and under voltage glitches, power starvation or overfeeding, temperature variations, light and laser shots to the CC, and clock glitches, have different mechanism of impacting a CC that is correlated with the precision (number of affected bits) and controllability (ability to reproduce the same fault) of the injected fault \cite{bao1997breaking, barenghi2010low, selmane2008practical, anderson1997low, aumuller2002fault, tobich2013voltage, fuhr2013fault, hutter2009contact,bar2006sorcerer, barenghi2012fault, skorobogatov2005semi}.
Voltage glitch attack (VGA) is a fault injection technique that exploits deliberately abrupt variation of the voltage level at the power supply of a CC \cite{hutter2009contact,bar2006sorcerer, barenghi2012fault, aumuller2002fault, tobich2013voltage}. 
Voltage glitches can lead to misinterpreting instructions of (crypto-) processor, failure to erase or overwrite data or retaining data from memory when not instructed~\cite{lee2015fault, weingart2000physical, skorobogatov2005semi}.

VGA is used in a fault injection attack on the RSA device in the presence of countermeasures in~\cite{aumuller2002fault}, and in~\cite{hutter2009contact} VGA is exerted to inject the faults in unprotected RFID tags.
A VGA attack on a Motorola MC68HC05B6 microcontroller by under-voltage attack is explained in \cite{skorobogatov2005semi} that modifies instruction of the microcontroller, and unauthorized access to the memory of a PIC16F84 microcontroller from Microchip using over-voltage glitches confirmed to be effective in the same literature.
Over-voltage VGA may burn-in the contents of CMOS random access memory (RAM) by failing the power down or overwrite instructions to erase the contents~\cite{weingart2000physical}.

Immense hardware countermeasures are proposed to further protect the cryptographic systems against side-channel attacks. 
Each of the countermeasures, even though created in general for one end, perseveres different aims. 
The utilization of digital methods, such as data encoding, masking, infective computation, and redundant computation, and the utilization of analog techniques, such as sensors, detectors, and noise addition are several of the available countermeasures against side-channel attacks.

Although countermeasures have been proposed to hinder fault injection attacks, most of these countermeasures require additional resources for CC that are not desirable.
Duplication and multiplication of CC, redoing processes, and coding data are types of countermeasures against fault attacks.
These methods, although possess a satisfactory immunity to fault attacks, have a significant overhead for power, performance, and availability of the CC. 
Consequently, fault injection resilient countermeasures, such as \cite{RSAfaultresilent}, have been proposed to provide security along with the availability. 

Various techniques have been proposed to counteract fault injection attacks.
Information redundancy-based techniques such as error correcting codes are a class of countermeasures against fault injection attack by encoding information flowing through the CC~\cite{barenghi2012fault}, while a recent paper \cite{bus_invert} provides a protectively against side-channel attacks using the bus-invert coding scheme. 
Spatial redundancy-based countermeasures are a class of countermeasures which use the duplication/multiplication of the hardware of the CC to ensure the accuracy of the output through majority voting, and temporal redundancy based countermeasures verify the output through repetition of (part of) the cryptographic algorithm in time.
Even though these countermeasures are advantageous in countering fault injection attacks, spatial, temporal, and information redundancies will lead to increased power dissipation of CC, reduced throughput, and increased area of the CC~\cite{barenghi2012fault, bar2006sorcerer,aumuller2002fault, aftabjahani2017robust}.
Alternatively, analog countermeasures, such as voltage, temperature, and frequency sensors, are used to detect malicious fault injection activities and to protect a CC by ceasing the operations if such an activity is detected~\cite{bar2006sorcerer,aumuller2002fault, beringuier2014voltage}.
Detection of dynamic supply voltage variations has been used in \cite{le2011long}.
Timing detectors are used to detect the glitches in \cite{beringuier2014voltage} as a digital solution to counteract the VGA within a specific voltage and clock range.
To the best of the knowledge of the authors, the on-chip VR has never been used as an inherent countermeasure against fault injection attacks.
This paper is the first work to utilize the existing resources of an on-chip VR as a countermeasure against voltage glitch attacks where the implications of on-chip VR and the number of phases are investigated in counteracting voltage glitch attacks.

To the best of the authors' knowledge, the benefits of the on-chip voltage regulator as an inherent countermeasure has not been discussed in the literature to thwart fault injection attacks. 
The hardware used in the voltage regulator intrinsically exists in the cryptographic circuits that are employed to prevent various side-channel attacks. 
However, the resistance of voltage regulators is not extended to voltage glitch fault injection attacks, while a fault injection attack can eliminate the protection of the circuit against side-channel attacks. 
In this paper, with an emphasis on the application of on-chip voltage regulator as an intrinsic defense mechanism to fault injection attacks, we confirm that the number of phases and capacitor size of the voltage regulator influences the immunity of a CC against voltage glitch attacks. 

Moreover, we introduce an infective computation based countermeasure that combines the on-chip voltage regulator with auxiliary fault detection and infection circuit. 
The combined countermeasure would try to lessen the influence of the injected glitch on the CC using the on-chip multiphase voltage regulator.
An infective countermeasure contaminates the cryptographic process using a pseudo-random number generator (PRNG) to make the leakages incompetent for an attacker when the MPVR is not adequate to prevent the injected glitch.
The proposed combined countermeasure increases the availability of the CC in the presence of the injected faults, while the protection of the voltage regulator as a countermeasure further improved using the infective computation in the presence of an on-chip voltage regulator as a countermeasure.
The proposed fault resilient countermeasure against voltage glitch attacks increases the availability of the CC in the presence of injected faults.

The rest of this paper is as follows.
In Section~\ref{sec:proposal}, the advantage of the on-chip VR on the resilience of CC against VGA is discussed and the effect of the capacitor size is analyzed.
In Section~\ref{sec:multiphaseVR}, the effect of increasing the number of phases in the robustness of the CC to the VGA is investigated.
In Section \ref{sec:simulations}, extensive practical evaluations on the S-box of an AES with and without an on-chip VR is presented, followed by discussions on the overhead of the proposed countermeasure and conclusions.

\section{Threat model}
For a fault injection attack, an attacker needs to inject a fault into the CC utilizing one of the techniques to generate faulty ciphertexts, and subsequently to interpret the observations using one of the fault analysis techniques to obtain the correct key. 
This section explains the threat model of a fault injection attack for the proposed countermeasure. 
It should be noted that not all the AES is simulated in this paper, and the explanations given in this section are to illuminate how a fault injection attack acts to obtain the key and to provide a vision for a well-designed countermeasure.
The offline fault analysis is out of the scope of this paper; however, knowledge for successful fault analysis is useful for designing effective countermeasures against fault injection attacks. 
The offline fault analysis techniques are succinctly explained to clarify the complete attack procedure.

\subsection{Fault injection}
\begin{figure}[]
\begin{center}
    \includegraphics[width=0.6\textwidth]{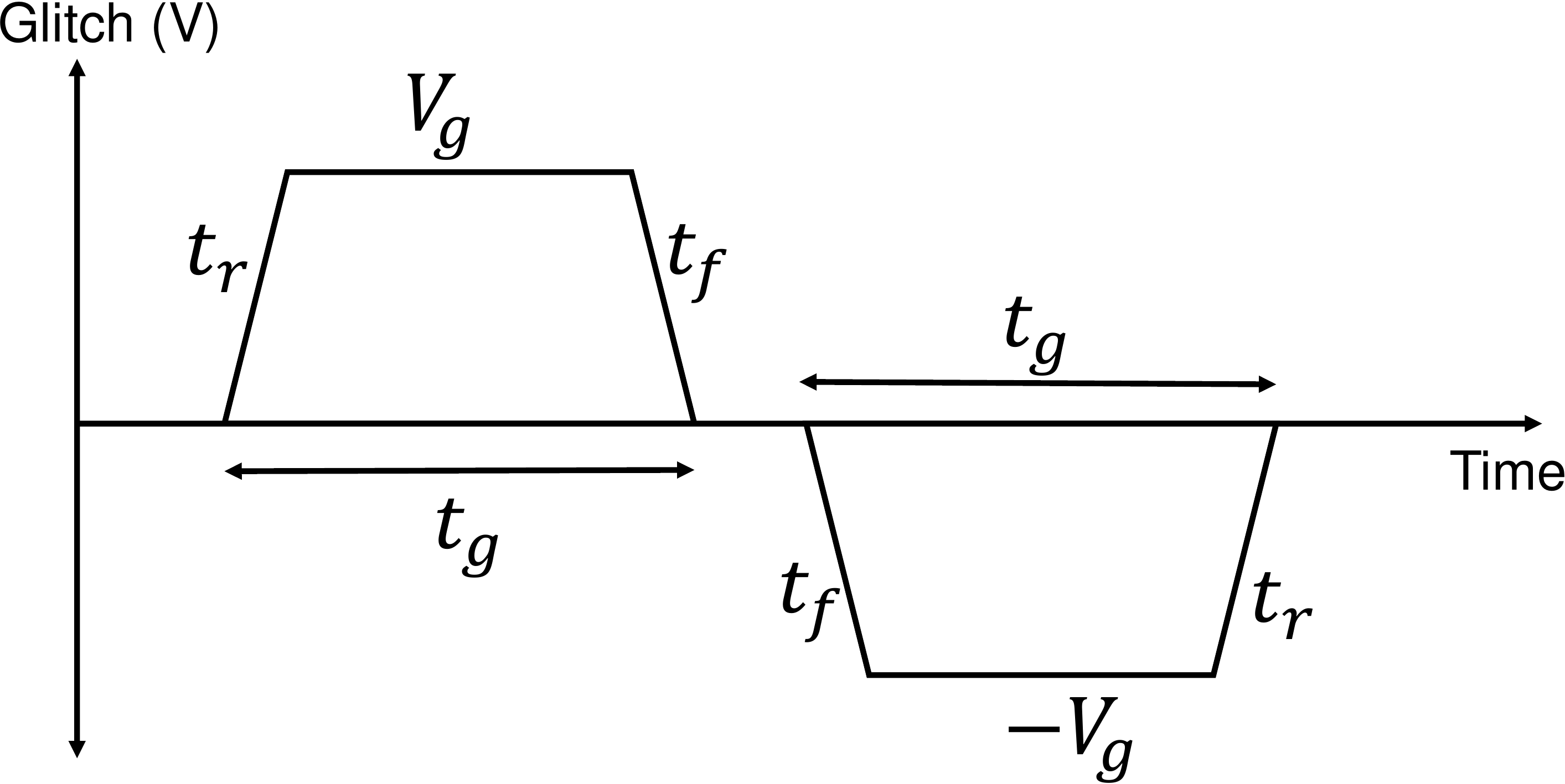}
\end{center}
\caption{A trapezoidal voltage glitch has a positive or negative amplitude depending on the choice of the attacker, and the rise time~$t_r$, fall time~$t_f$, and the duration~$t_g$ of the glitch.}
\label{fig:glitchdurationANDshape}
\end{figure}

The naive CC is assumed to have no countermeasure except the proposed countermeasure if needed. 
The attacker is equipped to perform a fault injection by varying the supply voltage of the CC to the desired level.
Moreover, the attacker is capable of injecting a voltage glitch into the CC at the desired time, with an arbitrary duration and amplitude. 
Fig. \ref{fig:glitchdurationANDshape} shows the glitch that the attacker applies to inject faults into the CC. 
The attacker craves to inject faults into the implementation of the CC.
Because the countermeasure is against the voltage glitch attacks, the assessment also comprises faults generated by voltage glitches. 
The proposed countermeasure is not evaluated to prevent different fault injection attacks such as temperature variations, clock glitches,  light, laser, and EM radiation attacks.
The attacker knows that the duration of the glitch should be at least half the length of the switching period of the CC, and assumes that the switching frequency of the CC is within 50 MHz and 1 GHz range.  
Similarly, the attacker comprehends that if the injected glitch voltage is too high for the CMOS device of the CC, the device will breakdown. 
Further, the attacker knows the voltage rating of the CMOS in CC and tries to inject glitches up to two times of the nominal voltage to infiltrate the CC in safe operation region. 
The attacker records the correct and faulty ciphertexts and inputs for an offline fault analysis phase

\subsection{Fault analysis}
Different techniques are introduced for fault analysis in fault injection attacks among which are differential fault analysis (DFA)~\cite{moradi2006generalized}, safe-error analysis (SEA)~\cite{yen2000checking}, and collision fault analysis (CFA)~\cite{blomer2006fault}. 
DFA is a technique that exploits correct-faulty pairs ciphertexts to analyze and obtain the secret key~\cite{moradi2006generalized}. 
In fault analysis using SEA, an attacker utilizes the leakages of the correct-faulty pairs of the ciphertexts for same input plaintexts~\cite{blomer2003fault}.
In CFA the collision at the output of CC for different input plaintexts is utilized to obtain the secret key~\cite{blomer2006fault}.

\begin{figure}[]
\begin{center}
    \includegraphics[width=\textwidth]{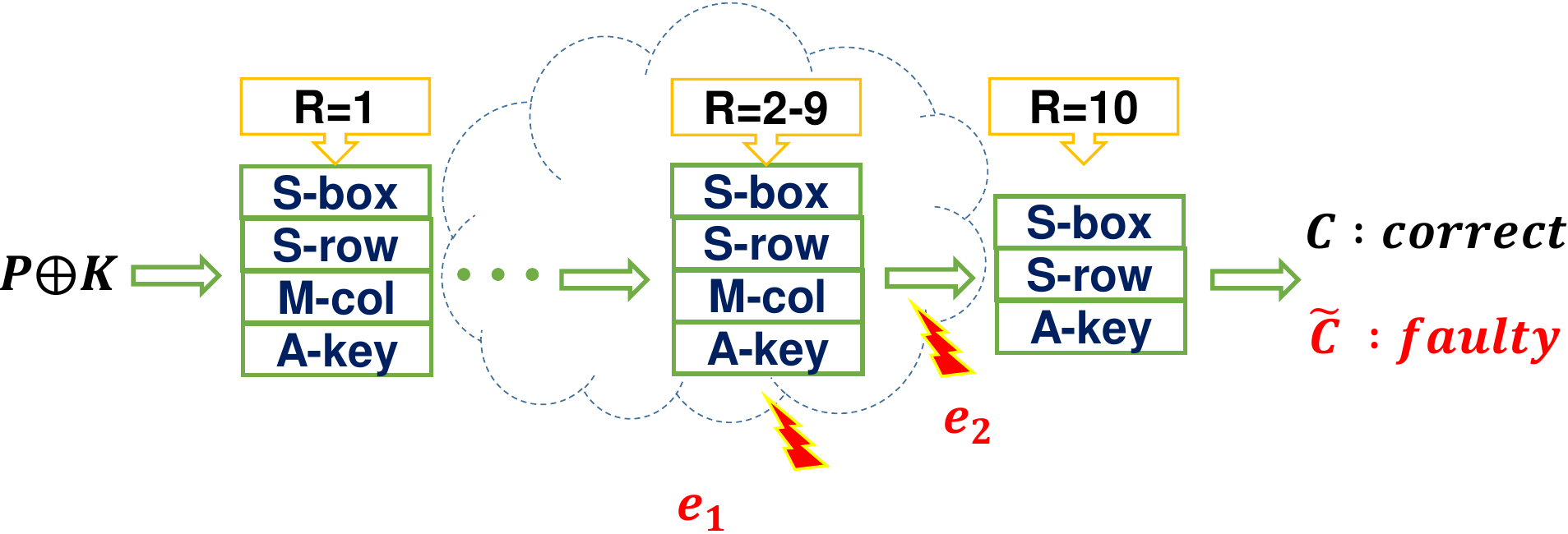}
\end{center}
\caption{Error $e_1$ occurs toward a byte of the block cipher in one of the rounds and diffuses as propagates in the CC, settling to error $e_2$ where  $e_1\neq e_2$.}
\label{fig:sp_network001}
\end{figure}

In Fig.~\ref{fig:sp_network001} an attacker injects a fault $e_1$ in the middle rounds of the AES cryptographic algorithm (with a substitution-permutation network), and $e_1$ is evolved to $e_2$ when reaches to the output where $e_1\neq e_2$ as the S-box is a nonlinear operation.
In~\cite{piret2003differential} a DFA on an AES is proposed that enables an attacker to obtain the correct key with only two pairs of correct-faulty ciphertexts using an injected fault in the $last$ or $(last-1)$ rounds of the CC.
In~\cite{moradi2006generalized} an extended version of~\cite{piret2003differential} is introduced by generalizing DFA to the injected fault within all rounds of an AES. 
Fault analysis of AES is proposed in~\cite{fuhr2013fault} that solely needs faulty ciphertexts without their correct pairs.

To perform an attack, an attacker selects a plaintext $\mathcal{P}$ and reaches the correct ciphertext $\mathcal{C}$.
Then, during the encryption initiates a voltage glitch in the supply of the CC.
Due to fault $e_1$ in Fig.~\ref{fig:sp_network001}, the correct state of the CC changes to the erroneous one $state' = state \oplus e_1$.
The error happens in one or multiple rounds or any part of the control circuitry of CC. 
Moreover the error may diffuse to more location of the CC due to inter-byte diffusion in the cryptographic algorithm~\cite{moradi2006generalized}, generating error $e_2$ (in Fig.\ref{fig:sp_network001}) which is nonlinearly related to $e_1$. 
Then the recorded faulty ciphertexts $\lbrace \Tilde{\mathcal{C}_1}, \Tilde{\mathcal{C}_2}, \Tilde{\mathcal{C}_3}, \dots \rbrace$ as explained in~\cite{moradi2006generalized, piret2003differential}.
The attacker repeats the experiment with different amplitudes and duration of voltage glitches or/and at a different time with the identical plaintext $\mathcal{P}$.
The correct-faulty pairs are used by the attacker to obtain the secret key in $\mathcal{DFA}(\mathcal{C},\lbrace \Tilde{\mathcal{C}_1}, \Tilde{\mathcal{C}_2}, \Tilde{\mathcal{C}_3}, \dots \rbrace )$; however, reproducibility and controllability of the errors are essential for a successful fault injection attack, \textit{i.e.}, the attacker has to be able to reproduce the fault for identical experiments (glitch timing, amplitude, duration).~\cite{roche2011combined, piret2003differential, moradi2006generalized}.

\section{On-chip VR as a countermeasure against VGA}\label{sec:proposal}
\begin{figure}[]
\begin{center}
    \includegraphics[width=0.6\textwidth]{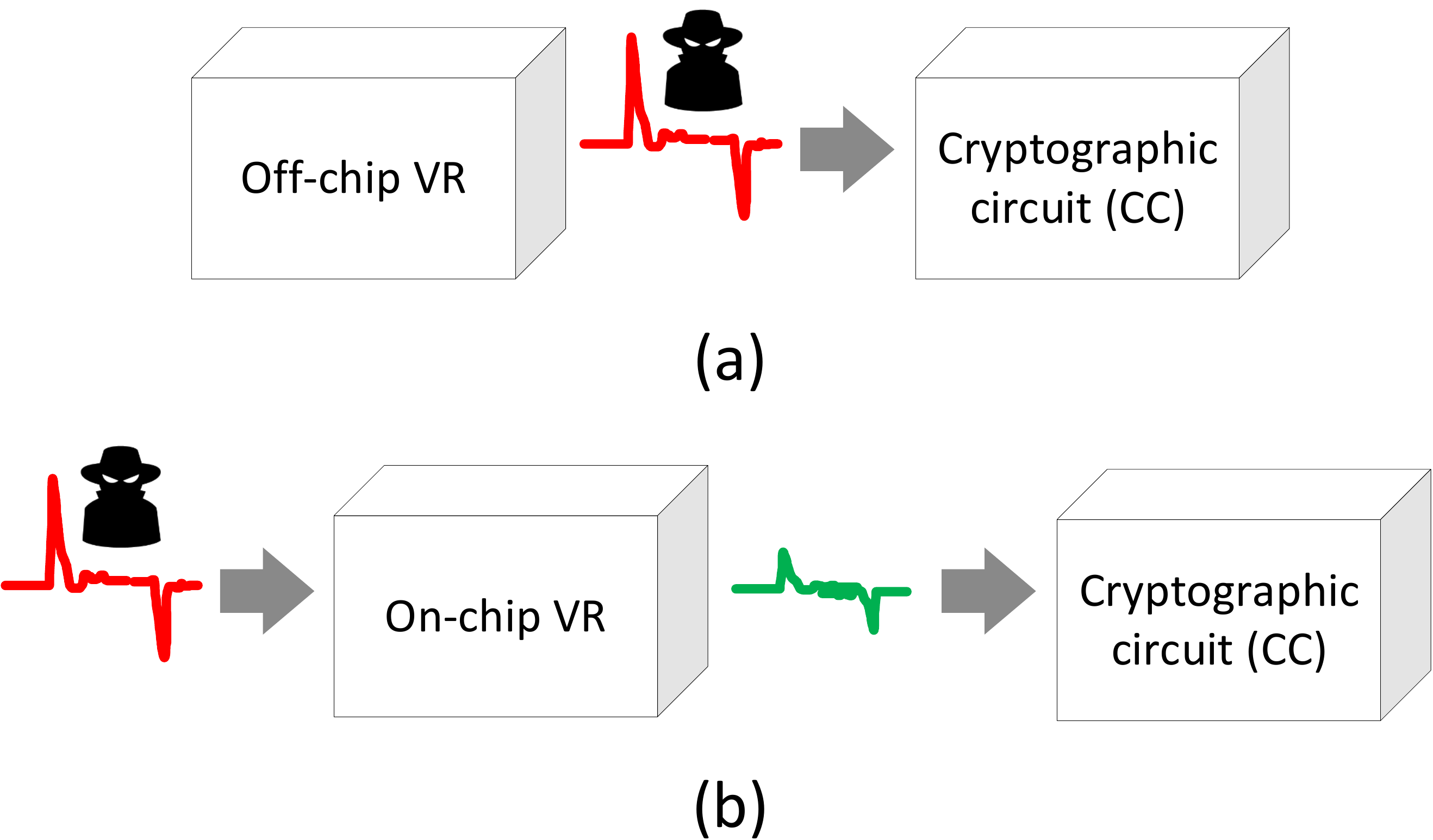}
\end{center}
\caption{A fault injection attack by a voltage glitch into a CC a) without on-chip VR and b) in the presence of an on-chip VR that debilitates the voltage glitches. }
\label{fig:on_chip_vs_off_chip_VR}
\end{figure}

In~\cite{yu2015leveraging, uzun2014converter, yu2016exploiting, khan2017implications, kar2016exploiting,kar20178} an on-chip VR is used as the first defense mechanism against power and EM attacks that motivates the use of a VR as a defense mechanism against fault injection by a power supply channel.
Implicitly VRs are utilized as triumphant countermeasures to tackle power and EM attacks; however, their vulnerabilities to fault attacks require an investigation.
\cite{vivek_isscc19} reports the susceptibility of the implemented on-chip VR-based countermeasure to fault injection attacks while the countermeasure is against the power and EM attacks. 
Different topologies of VR, such as LDO, buck, and switched-capacitor VR differently respond to the injected glitch at the input, and these differences affect how the glitches lead the CC. 
Fig.~\ref{fig:on_chip_vs_off_chip_VR} shows a voltage glitch injection to a CC in the presence of an on-chip VR and without on-chip VR. 
A subtle comparison of different topologies of the LDO, the buck, and the SC seems biased due to the differences in the components employed. 
Notwithstanding, the use of higher reactive components will increase the order of low-pass filtering in a VR. 
Hence, as in an LDO, a direct path is provided for the glitch to reach the CC, a buck VR will be considered to be more secure than an LDO against injected voltage glitches as the inductor limits the injected glitch in a buck.

\begin{figure}[]
\begin{center}
    \includegraphics[width=0.6\textwidth]{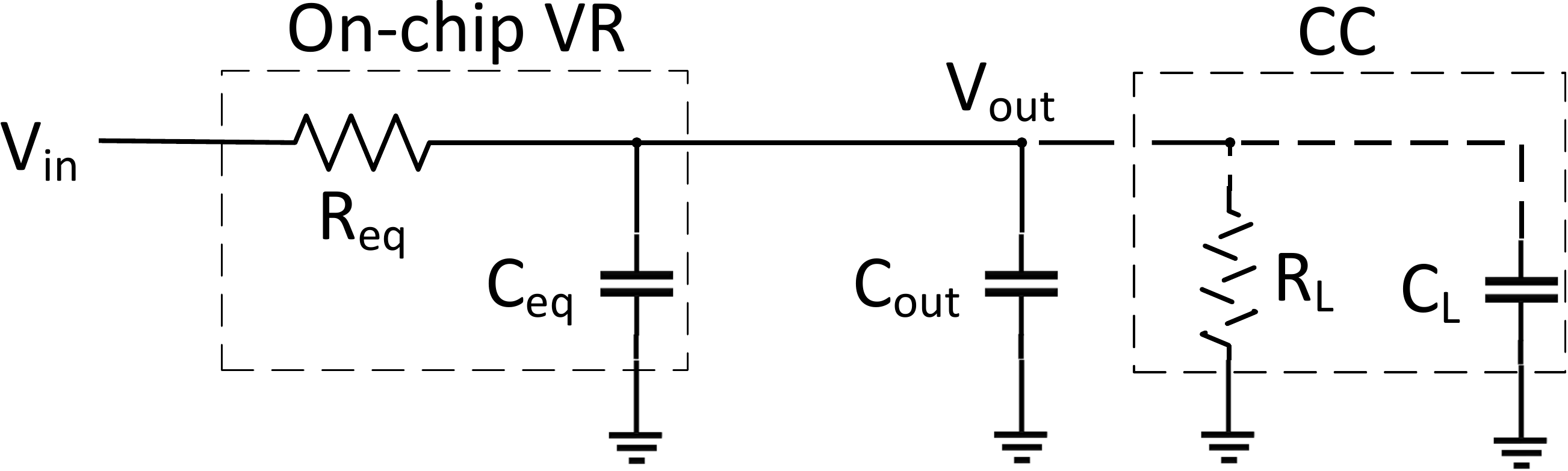}
\end{center}
\caption{A general, simplified first-order model for the behavior of a VR as a low-pass filter~(LPF) for input glitches. The impedance of the load (CC) comprises $R_L$ and $C_L$. $R_{eq}$ and $C_{eq}$ are representative of approximated first-order LPF of on-chip VR. }
\label{fig:VR_as_filter}
\end{figure}

\begin{figure}[]
\begin{center}
    \includegraphics[width=0.6\textwidth]{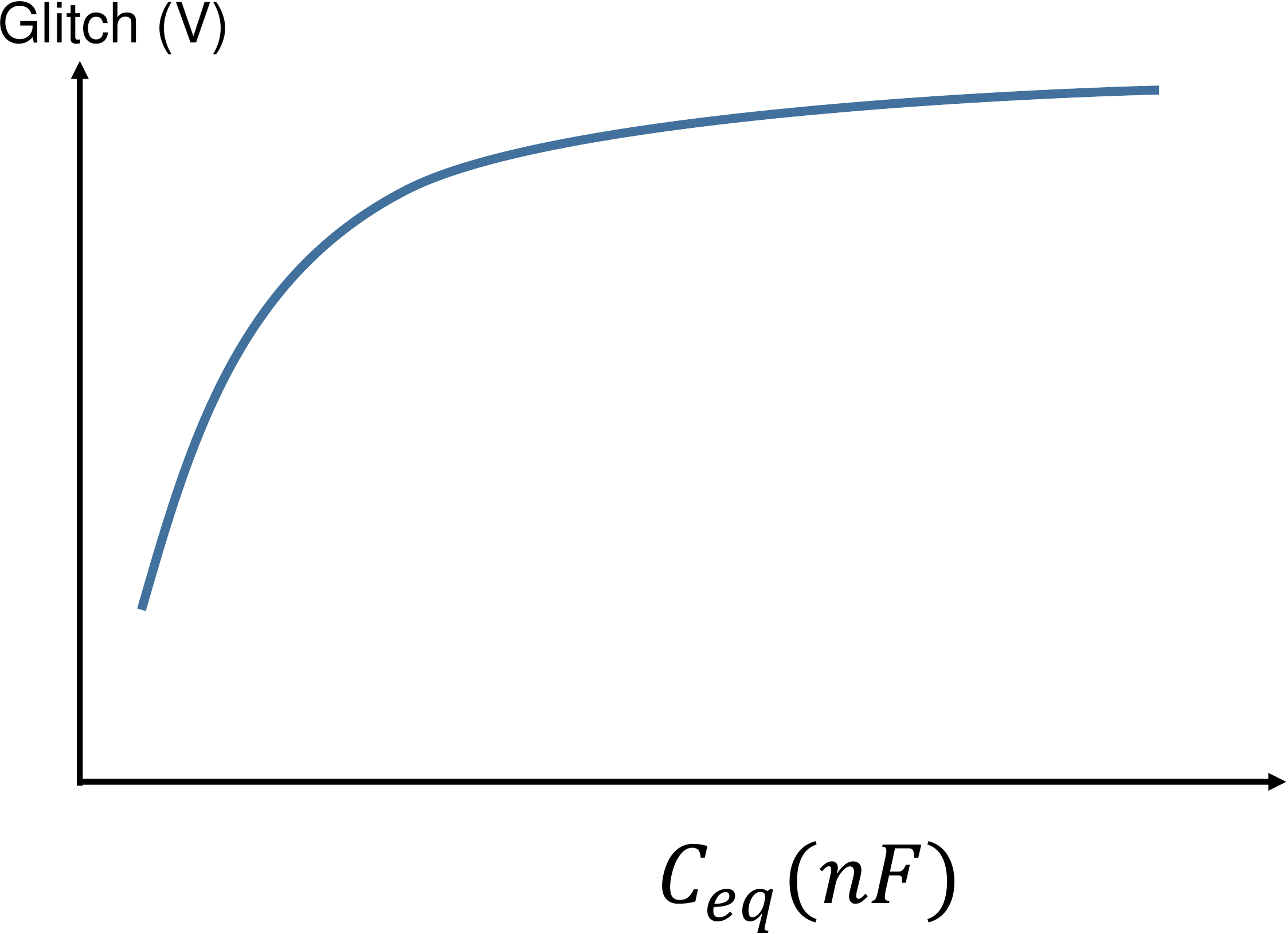}
\end{center}
\caption{Relation of capacitor size $C_{eq}$ and voltage glitch transmitted to the CC. With the increase of the $C_{eq}$ the voltage glitch transferred by the VR also increases to the extent that the cutoff frequency of VR appears, and after that, the increase in the amount of the capacitor does not have much effect on the transferred glitch energy. }
\label{fig:cap_vs_glitch_energy}
\end{figure}

In Fig.~\ref{fig:VR_as_filter} the first-order low-pass filter is depicted as a low-pass behavioral model of a VR. 
The number of reactive components in a VR can be one or more depending on the topology and configuration of the VR. 
The first-order model is not a perfect approximation of the response of a VR, but it can provide an insight into the behavior of the voltage converter by varying the size of the elements and the switching frequency. 
For instance, the latter analyzes the behavior of a switched-capacitor VR with the change of switching frequency and flying capacitance. 
Fast-switching limit (FSL) and slow-switching limit (SSL) introduces the optimum sizing of the components and the switching frequency of an SC-VR~\cite{seemanAnalysisOptimization}. 
However, the SSL and FSL describe the low-frequency (near DC frequencies) of the VR. 
By assuming $C_{eq}$ as the representative of the capacitor size, and $R_{eq}$ as the representative of the FSS and SSL, the following relationships can be written as
\begin{align}\label{eq:fsl_ssl_response}
&V_{out}(s)=\frac{1}{\frac{R_{eq}}{R_L}+sR_L(C_L+C_{out}+C_{eq})}Vin(s),
\end{align}
where
\begin{align}\label{eq:fsl_ssl_impedance}
&R_{eq}=\sqrt{R_{FSL}^2+R_{SSL}^2}.
\end{align}
By replcaing $R_{FSL}=\beta_{top} R_{on}$ and $R_{SSL}=\frac{{\gamma_{top}}}{C_{eq}f_{sw}}$~\cite{seemanAnalysisOptimization} into \eqref{eq:fsl_ssl_impedance}, the following expression could be achieved for \eqref{eq:fsl_ssl_response}
\begin{align}\label{eq:fsl_ssl_extended}
&V_{out}(s)=\frac{Vin(s)}{\frac{\sqrt{(\beta_{top} R_{on})^2+(\frac{{\gamma_{top}}}{C_{eq}f_{sw}})^2}}{R_L}+sR_L(C_L+C_{out}+C_{eq})}. 
\end{align}
When the VR is operating in an optimal region, $\frac{\sqrt{(\beta_{top} R_{on})^2+(\frac{{\gamma_{top}}}{C_{eq}f_{sw}})^2}}{R_L}\approx 1$, and given $s=j2\pi{f_{in}}$ the expression \eqref{eq:fsl_ssl_extended} is as follows
\begin{align}
&V_{out}(f_{in})=\frac{Vin(f_{in})}{1+j2\pi{f_{in}}R_L(C_L+C_{out}+C_{eq})},
\end{align}
Which $f_{in}$ is the frequency of the voltage glitch attack.
Therefore, the frequency response of the VR will be
\begin{align}
&\vert\frac{V_{out}(f_{in})}{V_{in}(f_{in})}\vert=\frac{1}{\sqrt{1+4\pi^2f_{in}^2R_L^2(C_L+C_{out}+C_{eq})^2}},
\end{align}
which affects the $f_{3dB}$ frequency of the VR.
Alternatively, the amount of energy transferred by the $C_{eq}$ to the CC is equal to $1/2C_{eq} (\Delta{V}_{C_{eq}})^2$, where $\Delta{V}_{C_{eq}}$ is the difference in voltage across the $C_{eq}$. 
Therefore, by increasing $C_{eq}$ the energy of the transferred glitch through the VR to the CC will be increased. 
Conversely, with increasing $C_{eq}$ the cutoff frequency $f_{3dB}$ of the VR reduces, thus reducing the energy of high-frequency VGA by the VR.
Therefore, there is a relationship between the capacitance of the VR and the glitch energy transmitted to the CC, as shown in Fig.~\ref{fig:cap_vs_glitch_energy}.

\section{Multi-phase VR against VGA}\label{sec:multiphaseVR}
For a switching circuit, the moment when a VGA occurs has implications on the success of the attack~\cite{djellid2006supply}. 
If the VGA occurs while the VR is connected to deliver the charge to the CC, the glitch does not reach to the CC and dissipates to the resistive off-switches.
Consequently, in a switching VR, the glitch strikes the CC if the attack occurs in a proper time. 
Therefore, switching VR degrades the likelihood of VGA success. 

\begin{figure}[]
\begin{center}
    \includegraphics[width=\textwidth]{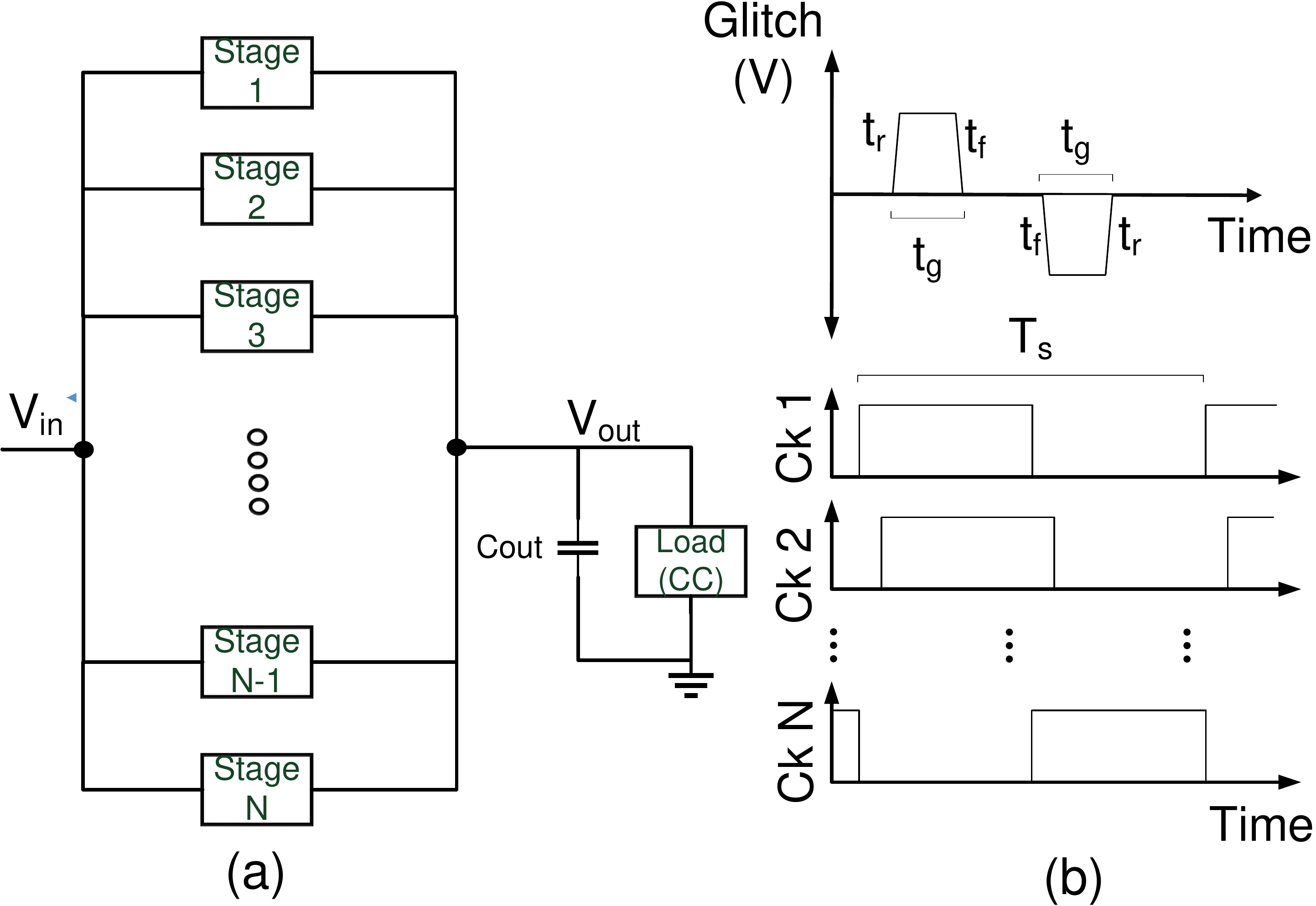}
\end{center}
\caption{Schematic of a) an MPVR with $N$ phases, and b) glitches of the VGA and clocks of the MPVR are depicted. The glitch has trapezoid shape with rise time $t_r$, fall time $t_f$ , and duration $t_{g}$.}
\label{fig:MPVR_concept}
\end{figure}

\begin{figure}[]
\begin{center}
    \includegraphics[width=0.6\textwidth]{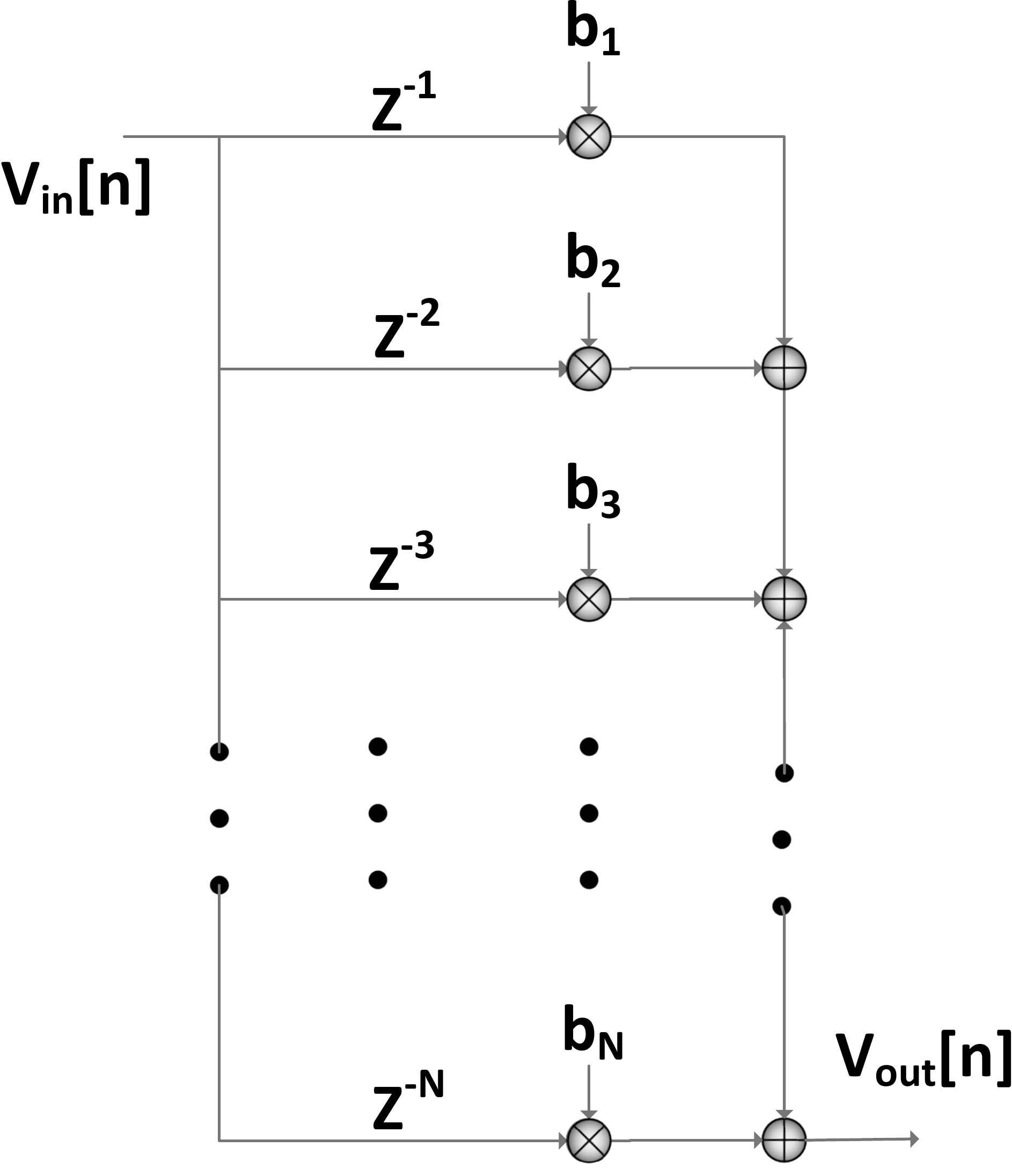}
\end{center}
\caption{An $N^{th}$ order FIR filter in which $V_{out}[n]=\sum_{i=1}^{N}b_iV_{in}[n-i]$.}
\label{fig:fir_filter}
\end{figure}

The block diagram of an MPVR with N interleaved stages (sub-converters) is shown in Fig.~\ref{fig:MPVR_concept}. 
MPVR is a technique in modern integrated circuits (IC) to enhance the performance of the VR in the generation, delivery, and management of power in modern electronic systems. 
The MPVR is a discrete time sampling system due to the timing that connects and disconnects continuously to and from the VR input. 
Due to the switching in a multiphase interleaved VR, and also because of the analogy of the MPVR with a finite-impulse-response (FIR) filter (shown in Fig.~\ref{fig:fir_filter}), interleaved VR exhibits attractive properties against VGA.  
Thus, an extension to consequences of the topology, component selection, and operating frequency, the number of phases in an MPVR determines the security of the CC against VGA.

\begin{figure}[]
\begin{center}
    \includegraphics[width=0.6\textwidth]{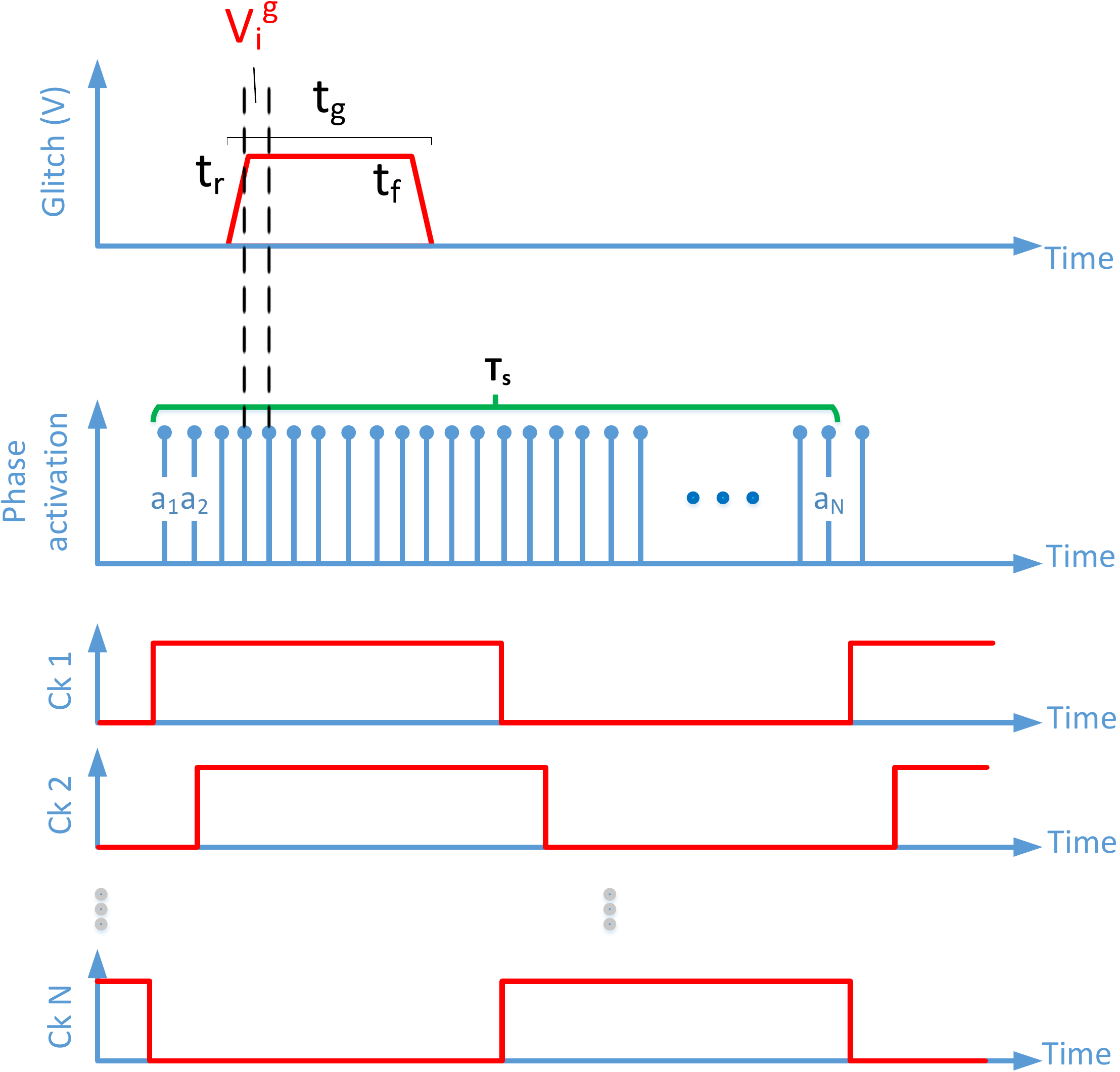}
\end{center}
\caption{Here is description of time domain glitch.}
\label{fig:time_domain_glitch}
\end{figure}

\begin{figure}[]
\begin{center}
    \includegraphics[width=0.6\textwidth]{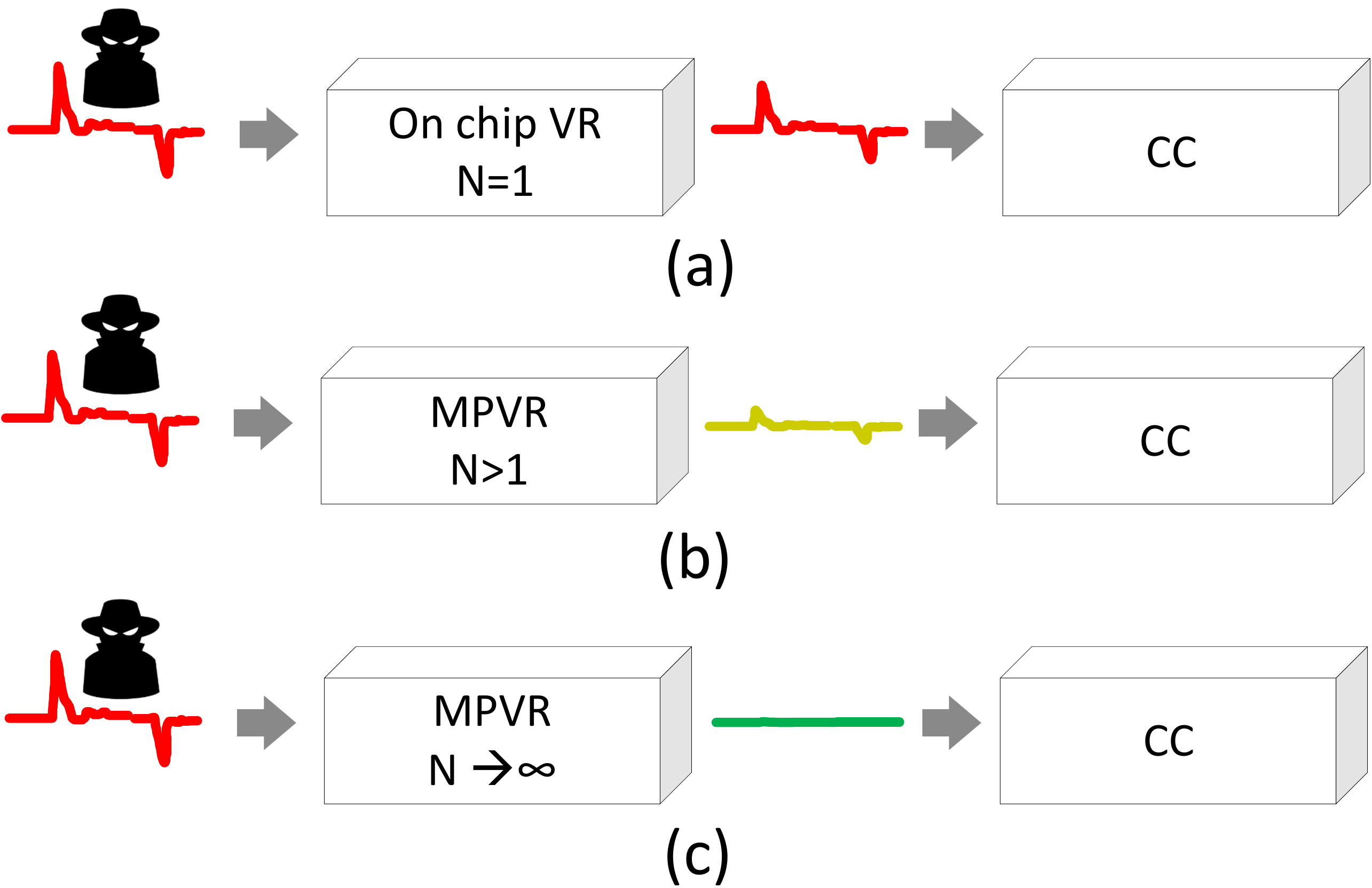}
\end{center}
\caption{Cryptographic circuit~(CC) is supplied by an on-chip voltage regulator. (a) For on-chip VR, VGA is healed a bit on CC, while for (b) $N>1$ the CC is less effected by VGA as compared to $N=1$ and for MPVR with (c) $N>>1$ the glitch is further reduced on the CC due to higher number of phases in on-chip VR.}
\label{fig:PVR_vs_CVR}
\end{figure}

The excessive energy spiking the CC during the VGA affects the success of the attack which is associated with the duration and amplitude of the input glitch~\cite{beringuier2014voltage}.
In an MPVR, by increasing the number of stages $N$ the transmitted power is reduced by each phase.
Considering the Fig.~\ref{fig:time_domain_glitch}, taking $V^g_i$ as the voltage glitch when the stage $i$ connects to the input of VR, the total glitch energy transmitted by the VR to the CC $E^g_{CC}$ at the end of period $T_s$ is as follows
\begin{align}\label{eq:ExcessEnergy}
&E^g_{CC}=\frac{C_{tot}}{2N}\sum\limits_{i=1}^{N} {(V^g_i)}^2,
\end{align}
where $C_{tot}=\sum^{}_{i}C_{fly,i}$ is the total fly capacitance of MPVR.
If the glitch duration is less than half of the $T_s$, such that $\frac{N}{2} > \Vert \overrightarrow{V^g}\Vert_0$, then
\begin{align}\label{eq:limit_of_glitch}
&\lim_{N\rightarrow \infty} E^g_{CC} \rightarrow 0.
\end{align}
\eqref{eq:limit_of_glitch} implies that as the number of the phases $N$ increases, the glitch on the CC reduces that results in reducing the effect of voltage glitch on CC, as shown in Fig.\ref{fig:PVR_vs_CVR}. 

\subsection{Frequency analysis of MPVR}\label{subsec:freqAnalysis}
\begin{figure}[]
\begin{center}
    \includegraphics[width=0.6\textwidth]{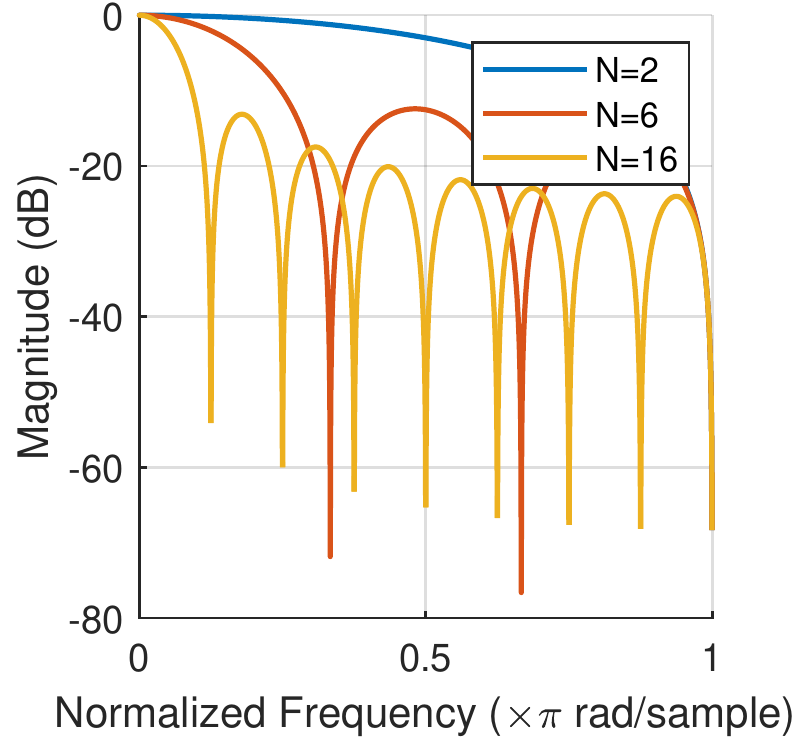}
\end{center}
\caption{FIR filter frequency behavior for different $N$ when $b_i$s are equal in Fig. \ref{fig:fir_filter} .}
\label{fig:FIR_for_various_N}
\end{figure}

According to the Nyquist condition, to reconstruct a signal from its samples, the sampling frequency should be at least twice the frequency of the sampled signal, and the reconstruction distorts the signal if the sampling outrages the condition. 
In this case, there is a relationship with the conditions obtained for the glitch duration (frequency) in \eqref{eq:limit_of_glitch} and the Nyquist condition for successful signal reconstruction. 
To achieve a countermeasure against a VGA using an on-chip VR the frequency of the VR has to be fewer than half the frequency of the injected glitch.
If this condition is fulfilled the injected voltage glitch on the CC is distorted by VR that makes it difficult for an attacker to establish a glitch.
Alternatively, by increasing the number of phases of an MPVR the order of the FIR filter also increases. 
Moreover, the coefficients $b_i$ are equal in Fig.~\ref{fig:fir_filter} for an MPVR.
Consequently, MPVR forms an FIR filter that its order increases with an increasing number of the phase $N$. 
Fig.~\ref{fig:FIR_for_various_N} shows the behavior of FIR filter for different $N$. 

\section{Fault-Resilient Voltage Regulator}
\begin{figure*}[h]
\begin{center}
    \includegraphics[width=\textwidth]{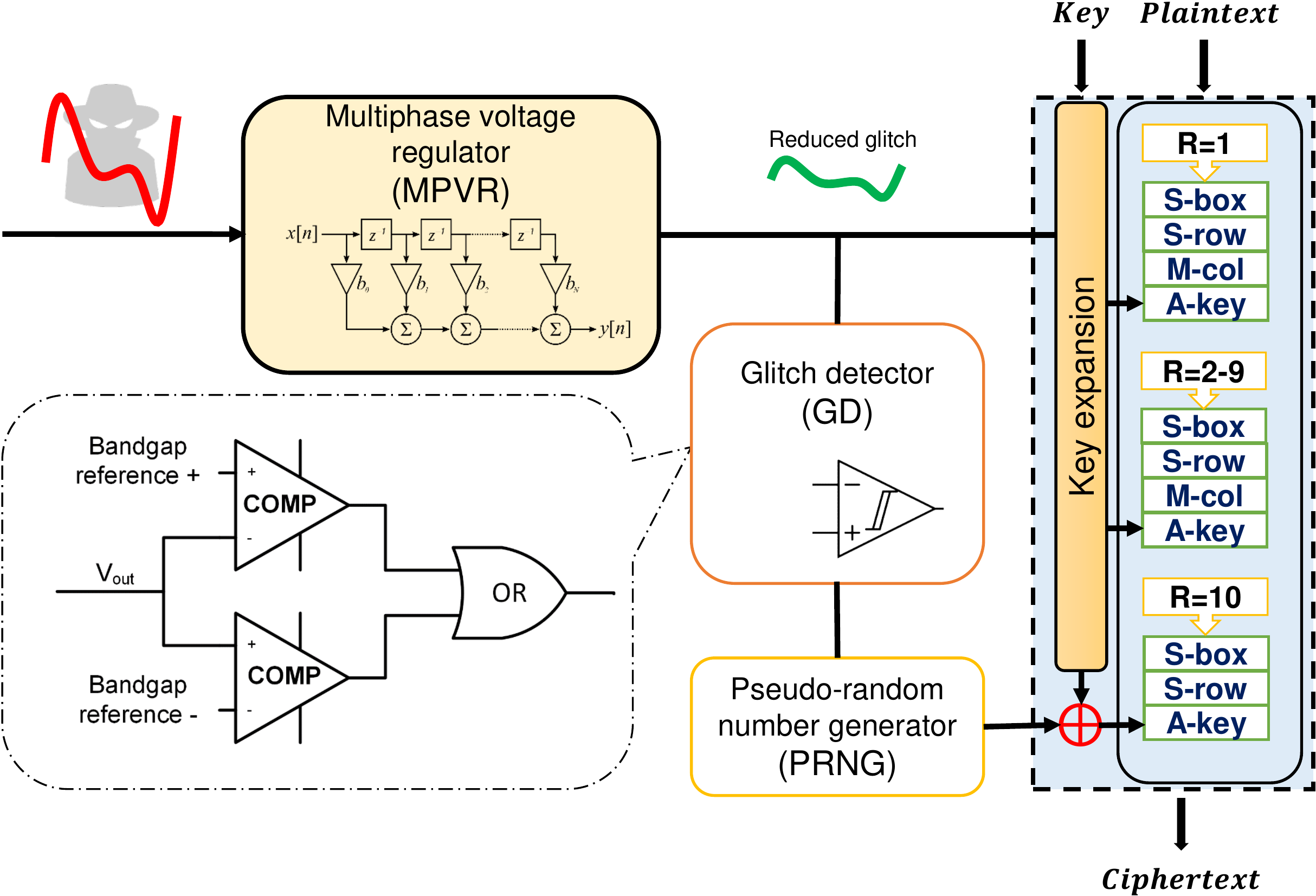}
\end{center}
\caption{The countermeasure is proposed for fault resiliency of the CC against voltage glitch attack. The MPVR strives to depreciate the significance of the injected glitch on the CC (an AES). In the case that the injected voltage glitch arrives at the CC for any reason, it is detected by the glitch detector (GD) circuit, and a random number is $XOR$ed with the key of the last round of the AES. In this case, the output of the operating CC depends on a random number (generated by PRNG) that makes the faulty ciphertext incompetent to an attacker. The GD is two comparators that compare the voltage glitch to the permissible voltages of the CC. The $OR$'ed output of the GD activates the PRNG to infect the computation.}
\label{fig:faultResilentVR}
\end{figure*}

Although a methodologically designed MPVR circuit reduces the energy of glitches on the CC, further advanced techniques for an attack such as combined fault attacks or voltage and clock glitches with frequency modifications is likely to affect the CC with a substantial glitch. 
Although the countermeasures against fault injection attacks, such as spatial, temporal, and information redundancies are remedial as a countermeasure, a defensive response of the countermeasure temporarily (or permanently) disables the CC if a fault attack occurs. 
Indeed, some of the countermeasures such as error correction codes can correct a specific number of erroneous bits while a fault occurs. 
Aforementioned suggests the intricacy of the availability of the CC to the safe use of the CC when a fault injection attack finishes.  
Our proposed technique for fault resilient countermeasure is based on the infective computation that if an stunning glitch concerns the CC, the countermeasure combines the correct key of the last round with a randomly generated number to mislead the attacker. 
Hence, the countermeasure randomly contaminates the output of the CC that dependents on the value of the random number and the glitch fault. 
Therefore, the output generated for the attacker will be futile. 
Fig.~\ref{fig:faultResilentVR} shows the architecture of the proposed countermeasure for an AES cryptography algorithm.

The negative and positive values of the reference voltage ranges are determined according to the tolerance of the CC and based on trial and error. This limit is related to static timing analysis of the CC.

\section{Practical evaluations}\label{sec:simulations}
\begin{figure}[]
\begin{center}
    \includegraphics[width=\textwidth]{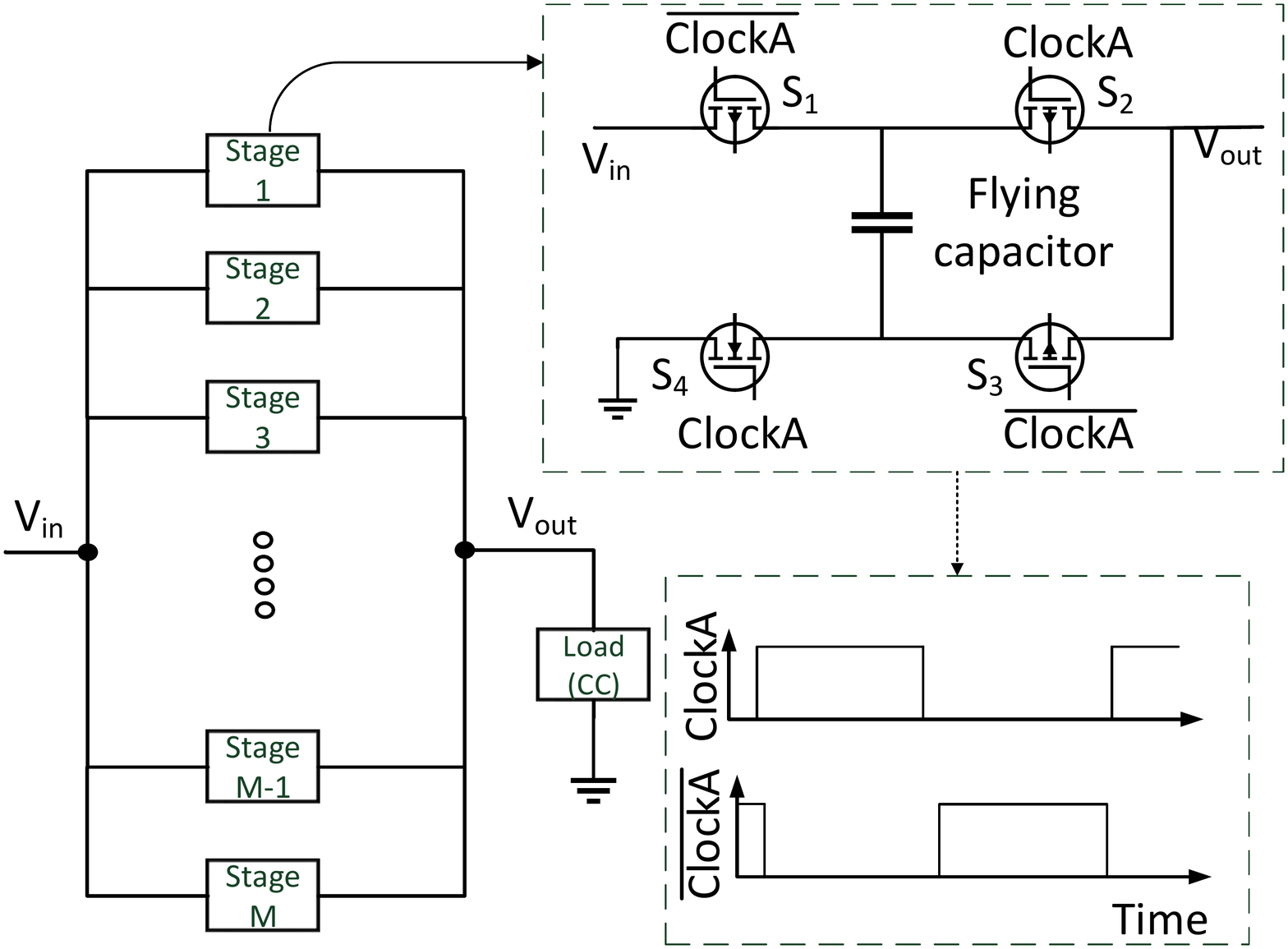}
\end{center}
\vspace{-0.1in}
\caption{A 2:1 SC-VR with $M$ stages is shown. Non-overlapping clocks A and B and switches $<S_1,S_3>$ are connected for $T_s/2 - \epsilon$, and switches $<S_2,S_4>$ are conducting for $T_s/2- \epsilon$ remaining, where $\epsilon$ is the time assigned to ensure non overlapping clocks.}
\label{fig:sc_core002}
\vspace{-0.2in}
\end{figure}

\begin{figure}[]
\begin{center}
    \includegraphics[width=0.6\textwidth]{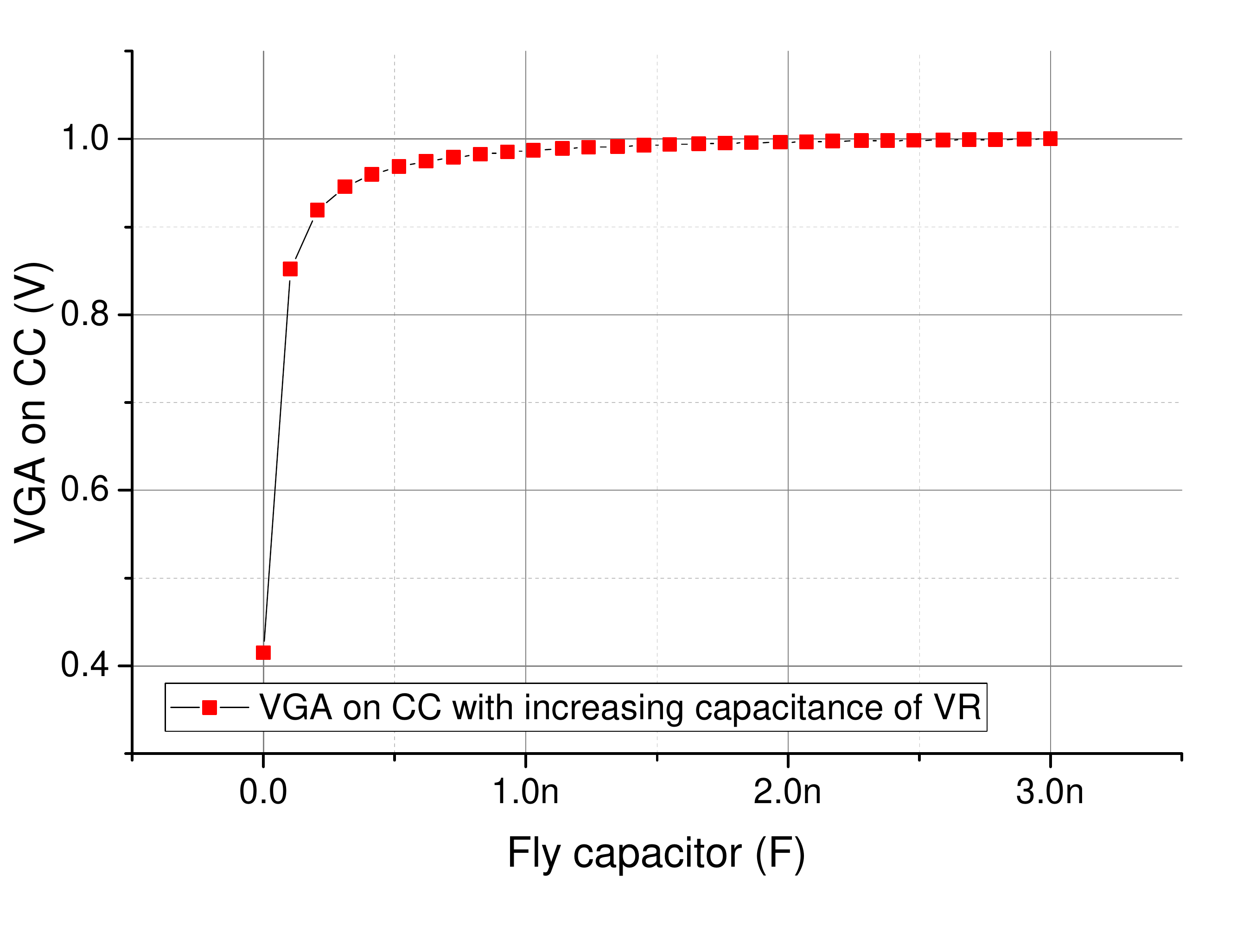}
\end{center}
\vspace{-0.3in}
\caption{The implication of the capacitive impedance of the on-chip VR on the effect of glitch on the CC. By increasing the size of the fly capacitor, the effect glitch on the CC increases.}
\label{fig:cap_size_vs_glitch}
\vspace{-0.1in}
\end{figure}

In this paper, SC-MPVR is preferred for simulations. SC-MPVR is a promising alternative across other counterparts, including the buck and LDO regulators due to the stability and  CMOS integration considerations~\cite{lu2018design}.
Moreover, an SC-MPVR with a high number of stages can be achieved by slicing larger capacitors and switches into tinier shares of capacitors and switches and utilizing an oscillator to form interleaved clock phases~\cite{uzun2014converter, lu2018design}.
A 2:1 SC-MPVR is designed and simulated in Virtuoso Cadence at $30-60MHz$~($T_s=16.67~nS$), $V_{in}=1.8V$, $V_{out}=0.9V$, and $M=\left\lbrace  1,\dots , 32 \right\rbrace$.
The schematic of the individual stages of VR, overall MPVR, and non-overlapping clocks are shown in Fig. \ref{fig:sc_core002}.
Switches $S_1,S_3$ are on for half a period and $S_2,S_4$ are on for the rest of $T_s$. 
For the optimization of the converter (switch sizing, capacitor size, frequency), the code referenced in \cite{seemanAnalysisOptimization} is used.
Moreover, an S-box of AES~\cite{yu2016exploiting} is implemented~\cite{ptm} using the Virtuoso Cadence to operate at $~200MHz$. 
Average power dissipation of the S-box is $256~\mu W$, where the minimum and maximum load power varies within $156.3$ and $387.22~\mu W$.  
As shown in Fig. \ref{fig:cap_size_vs_glitch} by increasing the size of the flying capacitor from $500~fF$ to $3~nF$, the effect of voltage glitch on the CC increases; however, after a certain point the raise is marginal due to the filtering behavior of the VR.

\begin{figure}[]
\begin{center}
    \includegraphics[width=0.6\textwidth]{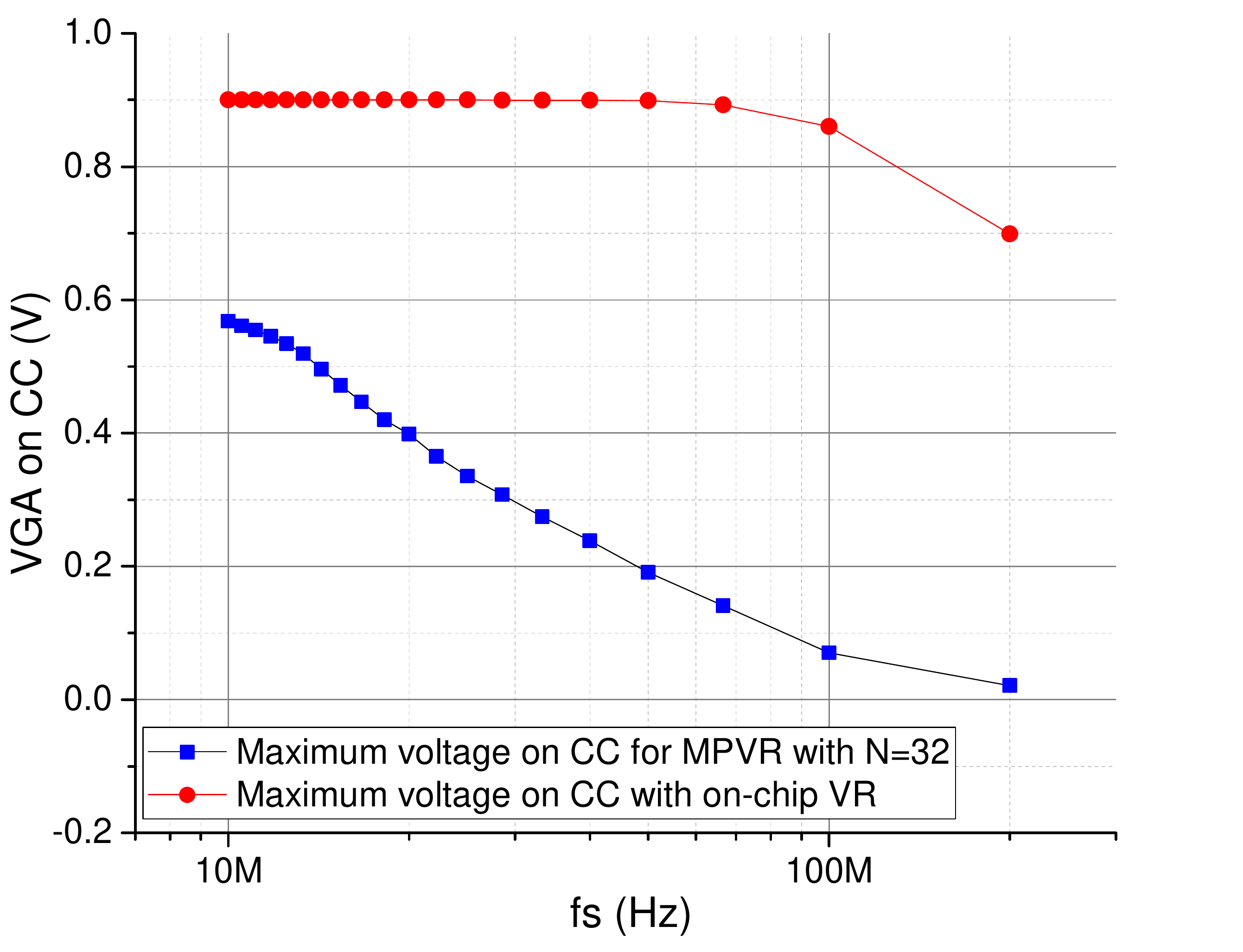}
\end{center}
\vspace{-.2in}
\caption{The maximum effect of VGA on a CC for different clock frequencies $f_{sw}$ of a VR is shown. For all frequency ranges, the resistance of MPVR with $N=32$~(depicted by squares) against VGA is higher than that of for a CC with on-chip VR~($N=1$).}
\label{fig:maxSpike_MP_vs_singlePhase}
\vspace{-0.1in}
\end{figure}

Fig. \ref{fig:maxSpike_MP_vs_singlePhase} shows effect of increasing the switching frequency $f_{sw}$ of the on-chip VR on the voltage glitch on the CC.
The resistance of the CC against VGA increases by increasing the switching frequency of the VR, and for all $f_{sw}$ frequencies, the robustness of the VR to VGA improves with increasing the number of phases.

\begin{figure}[]
\begin{center}
    \includegraphics[width=0.6\textwidth]{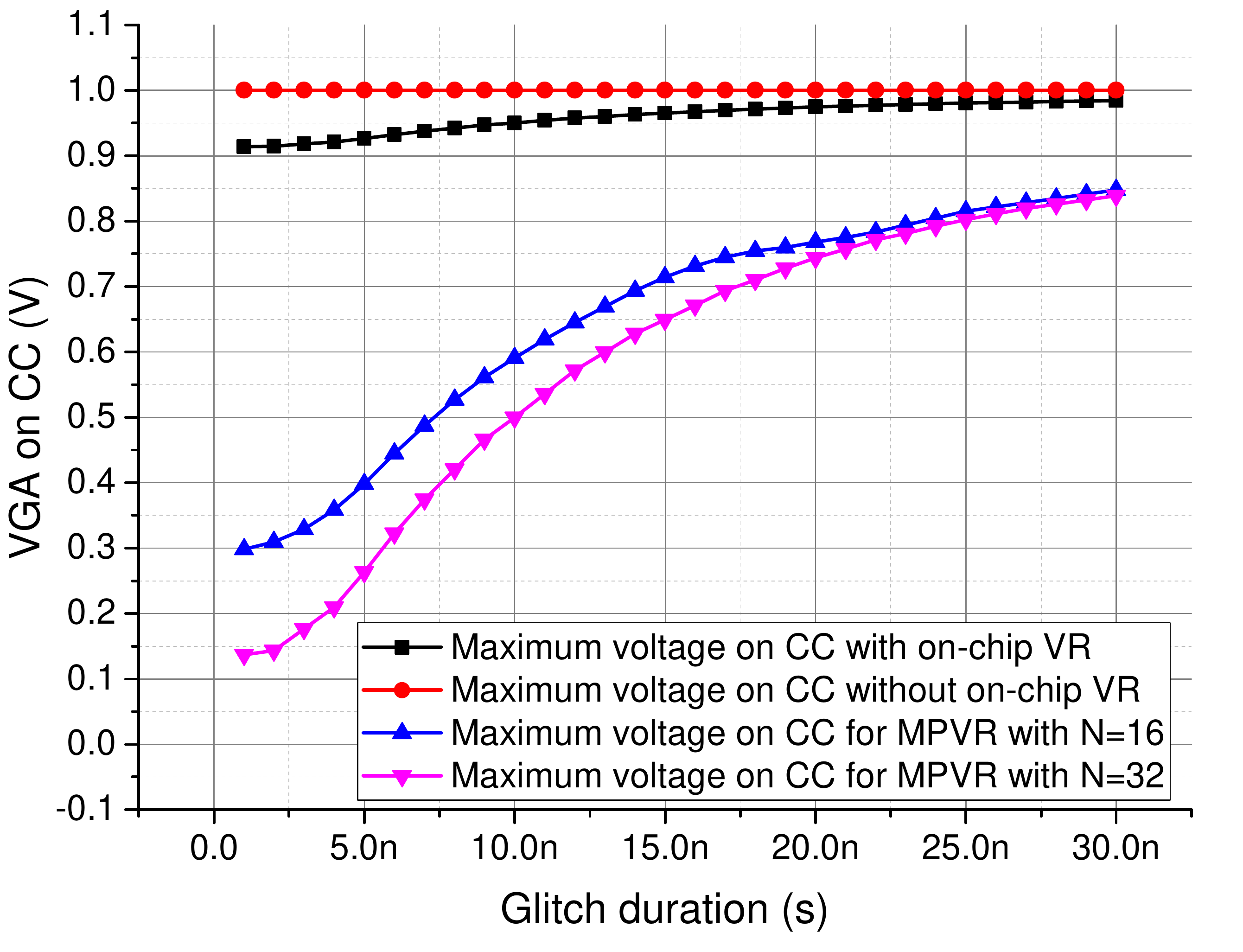}
\end{center}
\vspace{-.1in}
\caption{Relation of glitch duration and maximum effect of VGA on CC with glitch duration $\left\lbrace 1ns,3ns, \dots, 31ns\right\rbrace $, for CC with various MPVRs $N=\left\lbrace 1, 16, 32\right\rbrace $~(depicted by triangles and squares), and for CC without on-chip VR~(depicted by circles).}
\label{fig:Spike_duration_vs_voltage}
%\vspace{-.2in}
\end{figure}

As shown in Fig.~\ref{fig:Spike_duration_vs_voltage}, the security of the CC against VGA enhances by increasing the number of interleaved stages.
For a glitch with a 10~ns duration, the attenuation in the glitch amplitude is doubled with 32 interleaved phases, while the practical span of a voltage glitch on a CC is half the cycle of the operating frequency~\cite{djellid2006supply}.
The proposed countermeasure increases the resistance of the CC to a wide range of glitch span.
Considering the condition of \ref{subsec:freqAnalysis}, the VR is prepared to prevent the glitches with the duration less than $\frac{1}{2*30MHz}\approx 17~nS$.

\begin{figure}[]
\begin{center}
    \includegraphics[width=0.6\textwidth]{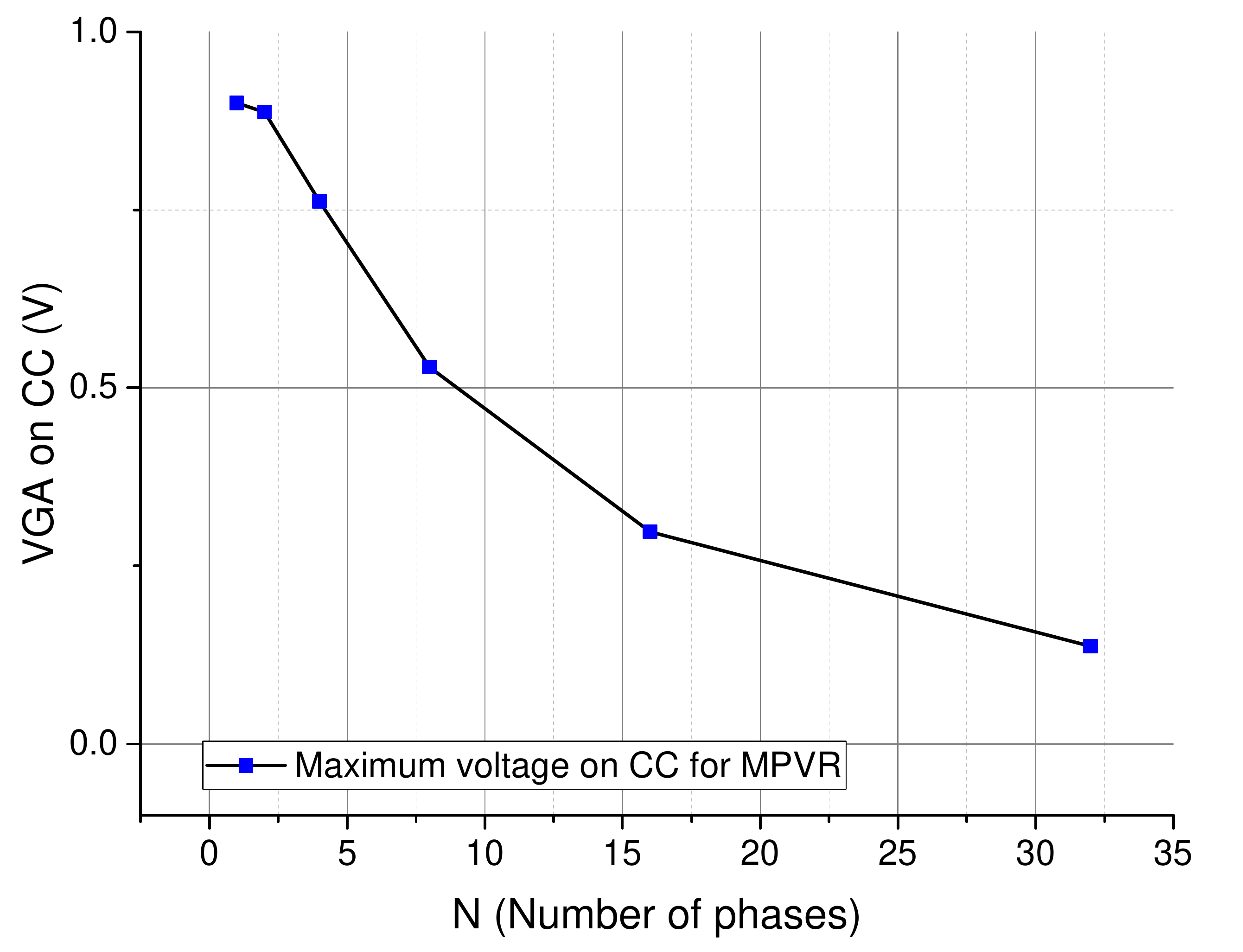}
\end{center}
\vspace{-.2in}
\caption{VGA on CC with on-chip VR for $M=\lbrace1, 2, 4, 8, 16, 32\rbrace$, when a glitch with amplitude $V_{g}=\pm 2$, duration $t_r=t_f=500ps$, and $t_{g}=1~ns$ is applied.}
\label{fig:number_of_phases_vs_glitch}
\end{figure}

\begin{figure}[]
\begin{center}
    \includegraphics[width=0.6\textwidth]{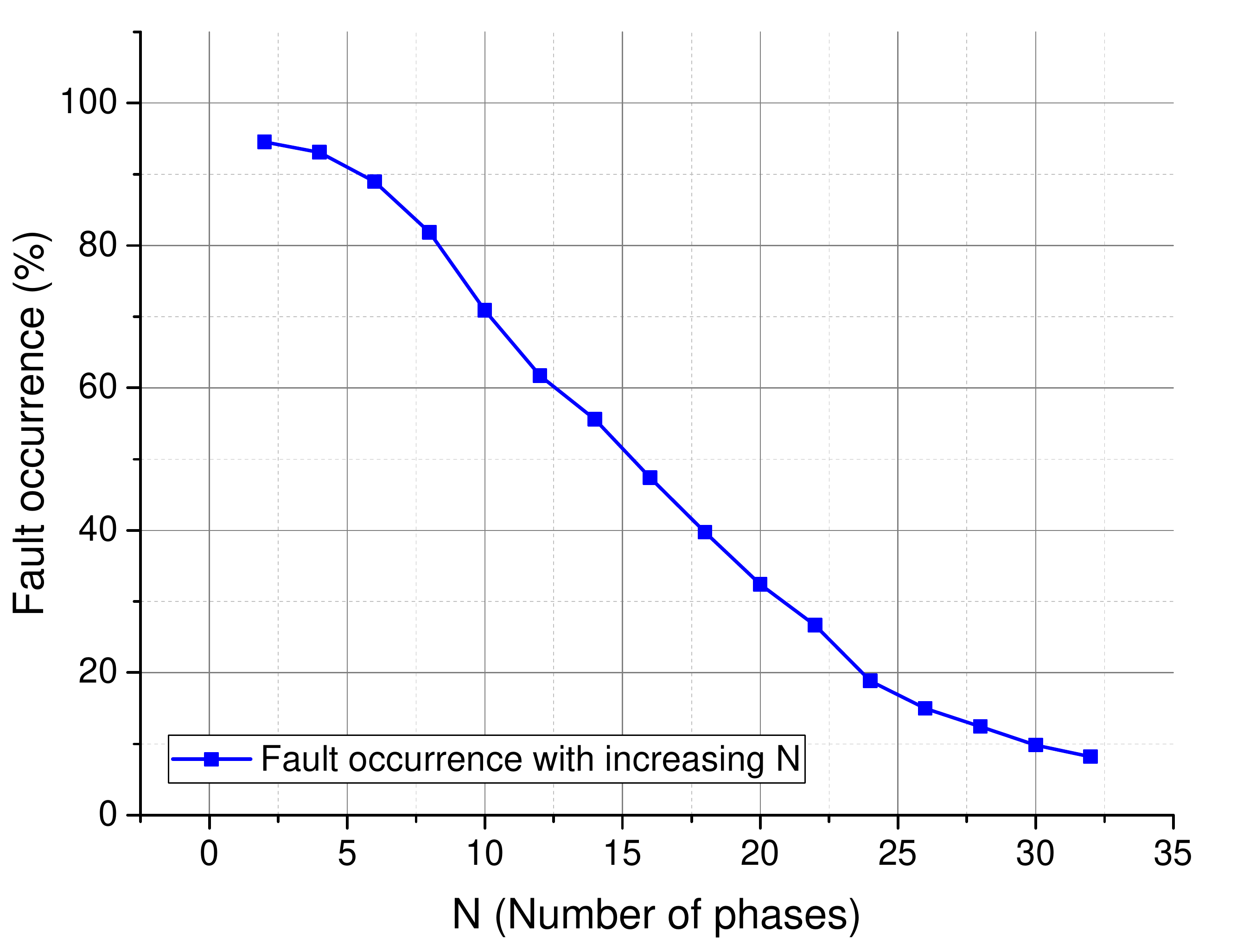}
\end{center}
\vspace{-.2in}
\caption{Fault occurrence versus the number of phases for an MPVR for a various number of phases is depicted. Glitch is injected to an S-box with MPVR and an S-box without MPVR that are compared to each other, and a counter specifies the number of faults at the output.}
\label{fig:SR_for_different_N}
\vspace{-.1in}
\end{figure}

As shown in Fig. \ref{fig:number_of_phases_vs_glitch},  the resistance of CC against VGA increases by increasing the number of phases $N$.
By specifying the faulty outputs as each result at the output of CC different than the correct one, the success of the VGA on the CC is determined by~\cite{selmane2008practical}
%
%\vspace{-.08in}
\begin{align}\label{eq:successFI}
&\%Glitch~attack~success=\frac{\#~Faults}{\#~All~tests}\times 100.
\end{align}
%\vspace{-.05in}
Using \eqref{eq:successFI} and repeatedly simulating the VGA on the S-box of an AES and counting the number of the faulty outputs using a comparator and a counter, the success rate of fault occurrence in the presence of an MPVR is obtained, as shown in Fig. \ref{fig:SR_for_different_N}.
The success of the VGA reduces by increasing the number of phases.
As can be seen in Fig. \ref{fig:SR_for_different_N}, while the fault coverage for the unprotected S-box is $0\%$, the fault coverage is $5.45\%$ with an on-chip VR, and the fault coverage reaches $51.45\%$ and $91.82\%$, respectively, with an increase in the number of phases to $16$ and $32$.
The fault coverage rate is assumed as one minus the fault rate.

\section{Discussion}
Assuming that the CC already has an on-chip VR, the throughput overhead on the CC is negligible. 
Although the increase in the number of phases of the VR is beneficial for the protection of a CC, by increasing the number of phases the losses in switches, buffers, and drivers increases~\cite{seemanAnalysisOptimization}.
Moreover, the design of an interleaved clock generators with higher resolution is an overhead in the design of MPVR with higher $N$.
The ripple at the output $V_{rip}$ is a function of output current, switching frequency of VR, the equivalent series resistance of capacitors of VR, and $N$, and decreases by increasing the number of phases~\cite{seemanAnalysisOptimization}.
Efficiency and area overheads of the on-chip VR with a various number of stages are listed in Table~\ref{tab:power}. 
To simulate the overhead values, the power, and area of the ring oscillator ring increases according to \cite{overhead}, while the power of the other elements varies with the number of phases according to~\cite{seemanAnalysisOptimization}.
The proposed countermeasure with MPVR can be useful for glitches with frequencies higher than twice the VR frequency.
\begin{table}[!t]
%\vspace{-.2in}
\caption{Efficiency and area overhead of MPVR.}\label{tab:power}
%\vspace{-.1in}
\begin{center}
\begin{tabular}{|l|l|l|l|l|l|l|l|}
\hline
M & 1 & 2 & 4 & 8 & 16 & 24 & 32\\ \hline
Ar.\% & 0 & 2.62 & 3.93 & 4.58 & 4.9 & 5.02 & 5.07 \\ \hline
Eff.\% & 84.4 & 84.54 & 84.68 & 84.9 & 85.56 & 86.0 & 85.41 \\ \hline
\end{tabular}
%\vspace{-.35in}
\end{center}
\end{table}
%

%\vspace{-.13in}
\section{Conclusion} \label{sec:conclusion}
In this paper, the application of an on-chip VR as a countermeasure against fault injection attack is proposed as a solution to enhance the robustness of the CC against a VGA.
The effect of the number of phases in the MPVR on the robustness of the circuit against VGA is analyzed.
The effectiveness of the proposed countermeasure on an S-box of an AES is evaluated.
The faults generated by the VGA on CC are reduced by $5.45\%$ with a single phase on-chip VR, and by $91.82\%$ with an MPVR with 32 phases, as compared to unprotected S-box of an AES device.
The throughput, power, and area overhead of the proposed technique are negligible due to the utilization of the existing VR as a power supply, while the area and power overhead of the MPVR are increased, respectively, by 5.1\% and 1\% when the number of interleaved phases is 32.
A voltage regulator as a countermeasure against fault injection attack has some limitations.
It is possible to generate a faulty output through the fault injection attack by increasing the duration or increasing the amplitude of the injected voltage glitch. 
To counteract this type of vulnerability to fault injection attacks, a new method based on infective computing has been proposed that contaminates data if a fault attack is detected. 
The combined countermeasure assures that if the glitch injection is successful, the attacker will not be able to obtain exploitable faulty outputs of the CC.

\section*{Acknowledgment}

%Dr. Reveryrand would like to acknowledge the funding by XLIM, Limoges, France. 
This work is supported in part by the NSF CAREER Award under Grant CCF-1350451, in part by the NSF Award under Grant CNS-1715286, in part by SRC Contract NO: 2017-TS-2773, and in part by the Cisco Systems Research Award. The authors would like to thank anonymous reviewers for their valuable inputs on improving this work.

% or


% ============================================
%\appendices
%Appendix one text goes here %\cite{Roberg2010}.

%\section{}

%\section*{Acknowledgment}

%The authors would like to thank D. Root for the loan of the SWAP. The SWAP that can ONLY be usefull in Boulder...

%\ifCLASSOPTIONcaptionsoff
%  \newpage
%\fi


%\begin{thebibliography}{1}
\bibliography{references}

\begin{thebibliography}{10}

\bibitem{vosoughi2019leveraging}
A.~Vosoughi and S.~K{\"o}se.
\newblock Leveraging on-chip voltage regulators against fault injection
  attacks.
\newblock In {\em Proceedings of the 2019 on Great Lakes Symposium on VLSI},
  pages 15--20, May 2019.

\bibitem{skorobogatov2016bumpy}
S.~Skorobogatov.
\newblock The bumpy road towards iphone 5c nand mirroring.
\newblock {\em arXiv preprint arXiv:1609.04327}, September 2016.

\bibitem{ali_iscas2019}
M.A. {Vosoughi} and S.~{K{\"o}se}.
\newblock Combined distinguishers to enhance the accuracy and success of side
  channel analysis.
\newblock In {\em IEEE International Symposium on Circuits and Systems}, pages
  1--5, May 2019.

\bibitem{moradi2006generalized}
A.~Moradi, M.~Shalmani, and M.~Salmasizadeh.
\newblock A generalized method of differential fault attack against aes
  cryptosystem.
\newblock {\em Cryptographic Hardware and Embedded Systems}, pages 91--100,
  October 2006.

\bibitem{piret2003differential}
G.~Piret and J.~J. Quisquater.
\newblock A differential fault attack technique against spn structures, with
  application to the aes and khazad.
\newblock {\em Cryptographic Hardware and Embedded Systems}, pages 77--88,
  September 2003.

\bibitem{giraud2004dfa}
C.~Giraud.
\newblock Dfa on aes.
\newblock In {\em International Conference on Advanced Encryption Standard},
  pages 27--41. Springer, May 2004.

\bibitem{dusart2003differential}
P.~Dusart, G.~Letourneux, and O.~Vivolo.
\newblock Differential fault analysis on aes.
\newblock In {\em International Conference on Applied Cryptography and Network
  Security}, pages 293--306. Springer, October 2003.

\bibitem{yen2000checking}
S.~M. Yen and M.~Joye.
\newblock Checking before output may not be enough against fault-based
  cryptanalysis.
\newblock {\em IEEE Transactions on Computers}, 49(9):967--970, 2000.

\bibitem{blomer2006fault}
J.~Bl{\"o}mer and V.~Krummel.
\newblock Fault based collision attacks on aes.
\newblock In {\em Fault Diagnosis and Tolerance in Cryptography}, pages
  106--120. Springer, October 2006.

\bibitem{barenghi2012fault}
A.~Barenghi, L.~Breveglieri, I.~Koren, and D.~Naccache.
\newblock Fault injection attacks on cryptographic devices: Theory, practice,
  and countermeasures.
\newblock {\em Proceedings of the IEEE}, 100(11):3056--3076, 2012.

\bibitem{bar2006sorcerer}
H.~Bar-El, H.~Choukri, D.~Naccache, M.~Tunstall, and C.~Whelan.
\newblock The sorcerer's apprentice guide to fault attacks.
\newblock {\em Proceedings of the IEEE}, 94(2):370--382, 2006.

\bibitem{bao1997breaking}
F.~Bao and et~al.
\newblock Breaking public key cryptosystems on tamper resistant devices in the
  presence of transient faults.
\newblock In {\em International Workshop on Security Protocols}, pages
  115--124. Springer, April 1997.

\bibitem{barenghi2010low}
A.~Barenghi and et~al.
\newblock Low voltage fault attacks to aes.
\newblock In {\em International Symposium on Hardware-Oriented Security and
  Trust}, pages 7--12. IEEE, June 2010.

\bibitem{selmane2008practical}
N.~Selmane, S.~Guilley, and J.~L. Danger.
\newblock Practical setup time violation attacks on aes.
\newblock In {\em Seventh European Dependable Computing Conference}, pages
  91--96. IEEE, May 2008.

\bibitem{anderson1997low}
R.~Anderson and M.~Kuhn.
\newblock Low cost attacks on tamper resistant devices.
\newblock In {\em International Workshop on Security Protocols}, pages
  125--136. Springer, April 1997.

\bibitem{aumuller2002fault}
C.~Aum{\"u}ller and et~al.
\newblock Fault attacks on rsa with crt: Concrete results and practical
  countermeasures.
\newblock In {\em International Workshop on Cryptographic Hardware and Embedded
  Systems}, pages 260--275. Springer, August 2002.

\bibitem{tobich2013voltage}
K.~Tobich and et~al.
\newblock Voltage spikes on the substrate to obtain timing faults.
\newblock In {\em Euromicro Conference on Digital System Design}, pages
  483--486. IEEE, March 2013.

\bibitem{fuhr2013fault}
T.~Fuhr, E.~Jaulmes, V.~Lomn{\'e}, and A.~Thillard.
\newblock Fault attacks on aes with faulty ciphertexts only.
\newblock In {\em Workshop onFault Diagnosis and Tolerance in Cryptography},
  pages 108--118. IEEE, August 2013.

\bibitem{hutter2009contact}
M.~Hutter, J.~M. Schmidt, and T.~Plos.
\newblock Contact-based fault injections and power analysis on rfid tags.
\newblock In {\em European Conference on Circuit Theory and Design}, pages
  409--412. IEEE, August 2009.

\bibitem{skorobogatov2005semi}
S.~P. Skorobogatov.
\newblock {\em Semi-Invasive Attacks: A New Approach to Hardware Security
  Analysis}.
\newblock PhD thesis, University of Cambridge, April 2005.

\bibitem{lee2015fault}
Y.~S.~Lee \textit{et al.}
\newblock Fault attacks by using voltage and temperature variations: An
  investigation and analysis of experimental environment.
\newblock In {\em Information Science and Applications}, pages 483--490.
  February 2015.

\bibitem{weingart2000physical}
S.~H. Weingart.
\newblock Physical security devices for computer subsystems: A survey of
  attacks and defenses.
\newblock In {\em International Workshop on Cryptographic Hardware and Embedded
  Systems}, pages 302--317, March 2008.

\bibitem{RSAfaultresilent}
Y.~Sung-Ming, S.~Kim, S.~Lim, and S.~Moon.
\newblock Rsa speedup with residue number system immune against hardware fault
  cryptanalysis.
\newblock In {\em International Conference on Information Security and
  Cryptology}, pages 397--413, December 2001.

\bibitem{bus_invert}
M.~A. {Vosoughi}, L.~{Wang}, and S.~{Köse}.
\newblock Bus-invert coding as a low-power countermeasure against correlation
  power analysis attack.
\newblock In {\em 2019 ACM/IEEE International Workshop on System Level
  Interconnect Prediction (SLIP)}, pages 1--5, June 2019.

\bibitem{aftabjahani2017robust}
S.~A. Aftabjahani and A.~Das.
\newblock Robust secure design by increasing the resilience of attack
  protection blocks.
\newblock In {\em International Verification and Security Workshop}, pages
  13--18. IEEE, July 2017.

\bibitem{beringuier2014voltage}
N.~Beringuier-Boher and et~al.
\newblock Voltage glitch attacks on mixed-signal systems.
\newblock In {\em Euromicro Conference on Digital System Design}, pages
  379--386. IEEE, August 2014.

\bibitem{le2011long}
H.~B. Le, X.~D. Do, S.~G. Lee, and S.~T. Ryu.
\newblock A long reset-time power-on reset circuit with brown-out detection
  capability.
\newblock {\em IEEE Transactions on Circuits and Systems II: Express Briefs},
  58(11):778--782, 2011.

\bibitem{blomer2003fault}
J.~Bl{\"o}mer and JP. Seifert.
\newblock Fault based cryptanalysis of the advanced encryption standard (aes).
\newblock In {\em International Conference on Financial Cryptography}, pages
  162--181, January 2003.

\bibitem{roche2011combined}
T.~Roche, V.~Lomn{\'e}, and K.~Khalfallah.
\newblock Combined fault and side-channel attack on protected implementations
  of aes.
\newblock {\em Smart Card Research and Advanced Applications}, pages 65--83,
  2011.

\bibitem{yu2015leveraging}
W.~Yu, O.~A. Uzun, and S.~K{\"o}se.
\newblock Leveraging on-chip voltage regulators as a countermeasure against
  side-channel attacks.
\newblock In {\em Design Automation Conference}, pages 1--6. IEEE, June 2015.

\bibitem{uzun2014converter}
O.~A. Uzun and S.~Kose.
\newblock Converter-gating: A power efficient and secure on-chip power delivery
  system.
\newblock {\em IEEE Journal on Emerging and Selected Topics in Circuits and
  Systems}, 4(2):169--179, 2014.

\bibitem{yu2016exploiting}
W.~Yu and S.~Kose.
\newblock Exploiting voltage regulators to enhance various power attack
  countermeasures.
\newblock {\em IEEE Transactions on Emerging Topics in Computing}, 2016.

\bibitem{khan2017implications}
A.~W. Khan, T.~Wanchoo, G.~Mumcu, and S.~K{\"o}se.
\newblock Implications of distributed on-chip power delivery on em side-channel
  attacks.
\newblock In {\em International Conference on Computer Design}, pages 329--336.
  IEEE, October 2017.

\bibitem{kar2016exploiting}
M.~Kar and \textit{et al.}
\newblock Exploiting fully integrated inductive voltage regulators to improve
  side channel resistance of encryption engines.
\newblock In {\em Proceedings of the 2016 International Symposium on Low Power
  Electronics and Design}, pages 130--135. ACM, August 2016.

\bibitem{kar20178}
M.~Kar and et~al.
\newblock 8.1 improved power-side-channel-attack resistance of an aes-128 core
  via a security-aware integrated buck voltage regulator.
\newblock In {\em International Solid-State Circuits Conference}, pages
  142--143. IEEE, February 2017.

\bibitem{vivek_isscc19}
A.~Singh and \textit{et al.}
\newblock 25.3 a 128b aes engine with higher resistance to power and
  electromagnetic side-channel attacks enabled by a security-aware integrated
  all-digital low-dropout regulator.
\newblock In {\em IEEE International Solid- State Circuits Conference}, pages
  404--406, February 2019.

\bibitem{seemanAnalysisOptimization}
M.~D. {Seeman} and S.~R. {Sanders}.
\newblock Analysis and optimization of switched-capacitor dc–dc converters.
\newblock {\em IEEE Transactions on Power Electronics}, 23(2):841--851, March
  2008.

\bibitem{djellid2006supply}
A.~Djellid-Ouar, G.~Cathebras, and F.~Bancel.
\newblock Supply voltage glitches effects on cmos circuits.
\newblock In {\em International Conference on Design and Test of Integrated
  Systems in Nanoscale Technology, 2006. DTIS 2006.}, pages 257--261, September
  2006.

\bibitem{lu2018design}
Y.~Lu, J.~Jiang, and W.~Ki.
\newblock Design considerations of distributed and centralized
  switched-capacitor converters for power supply on-chip.
\newblock {\em IEEE Journal of Emerging and Selected Topics in Power
  Electronics}, 6(2):515--525, 2018.

\bibitem{ptm}
NIMO Group Arizona~State University.
\newblock Predictive technology model (ptm), 2008. [Online]. Available:
  http://ptm.asu.edu/.

\bibitem{overhead}
Y.~{Lu}, J.~{Jiang}, and W.~{Ki}.
\newblock Design considerations of distributed and centralized
  switched-capacitor converters for power supply on-chip.
\newblock {\em IEEE Journal of Emerging and Selected Topics in Power
  Electronics}, 6(2):515--525, June 2018.

\end{thebibliography}
\bibliographystyle{unsrt}  

%\end{thebibliography}

\end{document}